\documentclass[journal]{IEEEtran}
\usepackage[T1]{fontenc}
\usepackage{cite}
\usepackage{amsmath,amssymb,amsfonts}
\usepackage{algorithmic}
\usepackage{graphicx}
\usepackage{textcomp}
\usepackage{pgfplots}
\usepackage{xcolor}
\usepackage{diagbox}
\usepackage[section]{placeins}
\usepackage{mathrsfs}
\usepackage{subfigure}
\usepackage{physics}
\usepackage{tikz} 
\usepackage{scalefnt}
\usepackage{qcircuit}
\usepackage{tabularx} 
\usepackage{enumerate}
\usepackage{multirow}
\usepackage{caption}
\usepackage{hyperref}

\usepackage{threeparttable}

\usepackage[ruled,vlined]{algorithm2e}

\newcommand{\leaveout}[1]{}
\newtheorem{example}{Example}
\newcommand{\RNum}[1]{\uppercase\expandafter{\romannumeral #1\relax}}

\begin{document}
\title{Quantum Circuit Transformation\\ Based on Tabu Search}
%
%
%

\author{Hui~Jiang, 
        Yuxin~Deng, 
        and~Ming~Xu
\thanks{H. Jiang, Y. Deng and M. Xu are with Shanghai Key Laboratory of Trustworthy Computing, 
	East China Normal University.}
}

%



\maketitle

\begin{abstract}
The goal of quantum circuit transformation is to  map a logical circuit to a physical device by inserting additional gates as few as possible in an acceptable amount of time. We present an effective approach called TSA to construct the mapping. It consists of two key steps: one makes use of a combined subgraph isomorphism and completion to initialize  some candidate mappings, the other dynamically modifies the mappings by using tabu search-based adjustment. Our experiments show that, compared with state-of-the-art methods GA, SABRE and FiDLS proposed in the literature, TSA can generate  mappings  with a smaller number of additional gates and it has a better scalability for large-scale circuits. 
\end{abstract}

\begin{IEEEkeywords}
Quantum circuit transformation,  Tabu search, subgraph isomorphism, initial mapping
\end{IEEEkeywords}

%
\IEEEpeerreviewmaketitle
\section{Introduction}
As we all know, quantum technology has been applied in practice.
However, the (great) improvements of computer science driven by quantum technology are still in the early stage,
since large quantum computers have not yet been built.
In 2017, IBM developed the first 5-qubit backend called IBM QX2, followed by the 16-qubit backend IBM QX3. The revised versions of them are called IBM QX4 and IBM QX5, respectively. IBM Q Experience~\cite{ibm} provides the public with free quantum computing resources on the cloud and opens source the quantum computing software framework Qiskit~\cite{qiskit}. 

Users of these early quantum computers mainly rely on quantum circuits to implement quantum algorithms.
There is a gap between the design and the implementation of a quantum algorithm~\cite{Almudever2020}. In the design stage, we usually do not consider any hardware connectivity constraints. But in order to implement an algorithm on a quantum  device, physical constraints have to be taken into account. For example, IBM physical devices only support 1-qubit gates and CNOT gates between two adjacent qubits. Hence, it is necessary to transform the circuits for quantum algorithms to satisfy both logical and physical constraints.  This process is called quantum circuit transformation, which maps a logical circuit to a physical device by inserting additional gates. A big challenge for quantum information is the problem of quantum decoherence. Due to the decoherence of qubits, quantum gates need to be applied in a coherent period as the time for a qubit to stay in a coherent state is very short. The longest coherence time of a superconducting quantum chip is still within 10--100us~\cite{2016Reagor}. 
Therefore, the main goal of quantum circuit transformation is to reduce the number of additional gates and the depth of output circuits in an efficient way.

In the current work, we shorten the lifetime of qubits by parallelization, and use IMSM~\cite{Sun2020} to generate partial isomorphic subgraphs of logical circuits 
and physical ones 
as part of the initial mapping. The advantage of the initial mapping is that we use the appropriate subgraph isomorphism and the two-way connection of the logical circuits and the physical ones to obtain a dense (clustered nodes) initial mapping, which avoids certain nodes from being mapped to remote locations. We use tabu search~\cite{Glover1990} to generate logical circuits that can be executed on the physical device. Our approach of quantum circuit transformation is thus called TSA.
Tabu search can avoid falling into local optima and swapping the recently swapped qubits, thereby improving the parallelism of quantum gates. We insert SWAP gates associated with the gates on the shortest path to the candidate set, which greatly reduces the search space and improves the search speed~\cite{Li2018}. We design three evaluation functions that consider not only  the current gates but also the constraints of the gates already processed.
Our experiment has been conducted by using the IBM Q20 architecture as the target physical device.
The experimental results show that the evaluation function based on calculating the number of additional gates inserts the fewest additional gates.
We test several combinations of state-of-the-art initial mapping and adjustment algorithms aiming to insert fewer additional gates after quantum circuit transformation. Generally speaking, TSA outperforms GA~\cite{Zulehner2017}, SABRE~\cite{Li2018} and FiDLS~\cite{2020Qubit} in different aspects. When compared with DLH~\cite{2020Zhu} which consists of two evaluation functions MCPE and MCPE\_OP, TSA  performs better on large-scale circuits on the DLH benchmarks.

The main contributions of this paper are summarized as follows.
	\begin{enumerate}
	\item 
    We extend IMSM, which only generates part of initial mappings, by completing the mapping based on the connectivity between qubits.
	\item We propose a heuristic circuit adjustment algorithm based on tabu search, which can adjust large-scale circuits much more efficiently than existing precise search and heuristic algorithms.
	\item  We propose three look-ahead evaluation functions for the circuit adjustment; one employs configuration checking with aspiration (CCA)~\cite{Cai2012}, and the other two use the number of additional gates and the depth of the generated circuit as evaluation criteria, taking into account both the current gates and some gates yet to be processed. 
	\item We compare several state-of-the-art initial mapping and adjustment algorithms, and the results show that the initial mapping generated by our method requires to insert fewer SWAP gates, and TSA has a better scalability than them
	for adjusting the mapping for large-scale circuits.
	\end{enumerate}

The rest of this article is organized as follows.
In Section~\ref{Related work} we discuss some related work. In Section~\ref{Background}
we recall some background of quantum computing and quantum information. In Section~\ref{Quantum Circuit Transformation} 
we introduce the problem of  quantum circuit transformation and provide our detailed solution.
The experimental results are reported in Section~\ref{Experiment}. 
We conclude in the last section and discuss some future work.

\section{Related work}
\label{Related work}
There exist several initial mapping methods. Paler~\cite{Paler2018} has shown that  initial mapping have an important impact on quantum circuit transformation. He has proposed a heuristic method to find the initial mapping. Just by placing qubits in different positions from the default trivial placement
 in the  circuit instances on  actual NISQ devices, the cost can be reduced by up to 10\%. Li et al.~\cite{Li2018} have proposed a novel reverse traversal technique, which determines the initial mapping by considering the entire circuit. Zhou et al.~\cite{Xiangzhen2020} have put forward an annealing algorithm to find an initial mapping, but it is unstable. In~\cite{2020Qubit}, Li et al. have considered the subgraph isomorphism algorithm FiDLS to generate an initial mapping.  Zhu et al.~\cite{2020Zhu} have proposed an expansion-from-center scheme to determine the initial mapping. Starting from the center of the interaction graph, they arrange all logical qubits in the order defined by breadth-first search (BFS), and explore all neighboring nodes at the current depth in a strict chronological order. The first mapped node of this method has an important impact on the entire initial mapping. If the relationship between the logical interaction graph and the coupling graph is not considered, the mapping from the center will lead to more additional gates to be inserted.

One important goal of circuit adjustment algorithms is to minimize the number of additional gates. There are currently five main methods to solve the quantum circuit adjustment problem.
\begin{itemize}
  \item
\emph{Unitary matrix decomposition algorithm.} It is used in~\cite{2019CNOT,2019Quantum} to rearrange a quantum circuit from the beginning while retaining the input circuit. It can be applied to a broad class of circuits consisting of generic gate sets, but the results are not as efficient as a compiler designed specifically for this task.
\item
\emph{Converting into some existing problems.} This approach converts the quantum circuit transformation problem into some existing problems, such as AI planning~\cite{2017Temporal,2018Integer}, integer linear programming~\cite{2019Almeida} and satisfiability modulo theories (SMT)~\cite{2019Murali}, and then uses existing tools to find the optimal results in an acceptable amount of time for the problem.
Furthermore, as the time cost is usually long, it can only process small-scale quantum circuits. 
\item
\emph{Exact methods.}
Siraichi et al. have proposed an exact method~\cite{2018QubitSiraichi}. It will iterate over all possible mappings for all dependencies, so it is only suitable for simple quantum coupling graphs and cannot be extended to complex ones.
\item
\emph{Graph theory.} 
In~\cite{Shafaei2013}, Shafaei et al. have used the minimum linear permutation solution in graph theory to model the problem of reducing the interaction distance. The main idea is to divide a given circuit into several sub-circuits and apply the minimum linear permutation solution, respectively. Then, by inserting additional gates, all gates in the sub-circuits can be executed. Finally, a bubble sort is used to calculate the number of inserted SWAP gates. In~\cite{Guerreschi2018,Matsuo2019}, a two-step method is used to reduce the quantum circuit transformation to a graph problem to minimize the number of additional gates, based on the graph coloring problem and the largest subgraph isomorphism problem.
\item
\emph{Heuristic search.}
Shafaei et al. have used reversal,  bridge or swap to achieve quantum circuit transformation~\cite{2018QubitSiraichi}. 
Zulehner et al. \cite{Zulehner2017} have suggested to layer the circuits, then determine compatible mappings for each of these layers to insert as few additional gates as possible.
Cowtan et al.~\cite{Cowtan2019} have given  a qubit routing method $t\ket{ket}$, which
defines a distance vector to approximate the number of SWAP gates in order to compute a sequence of sets of candidate SWAPs. 
Li et al.~\cite{Li2018} have proposed a SWAP-based search algorithm called SABRE. 
It uses a heuristic evaluation function that trades off the number of 2-qubit gates and the depth of the circuit. Compared with previous search algorithms based on exhaustive mapping, SABRE can handle large-scale quantum circuits. But it depends on a random initial mapping, which does not seem to be the best choice in our opinion. 
Zhou et al.~\cite{Xiangzhen2020} have designed a heuristic search algorithm with a novel selection mechanism. Instead of choosing the operation with the lowest cost to apply, one can look ahead one step and then choose the best continuous operation. In this way, the algorithm can effectively avoid local optima. Moreover, a pruning mechanism has been introduced to reduce the size of search space and ensure that the program terminates in a reasonable amount of time. Li et al.~\cite{2020Qubit} have suggested to use filtered  depth-limit search  and feedback to minimize the number of SWAP gates. Their method, FiDLS, tends to search through all possible combinations of SWAP gates to maximize the number of executable 2-qubit gates. But the cost of a thorough search is very high, thus it sets a fixed limit and “filters” out those SWAP gates that do not interact with the gates in the front layer of the circuit.  Using exhaustive search, the number of auxiliary gates introduced by FiDLS is small, but the time cost is very high, especially when dealing with medium-scale and large-scale circuits. 
Zhu et al.~\cite{2020Zhu} have put forward a dynamical look-ahead heuristic cost function to adjust the window size according to the details of the quantum circuit. With the support of the dynamic look-ahead technique, they can deal with some large-scale benchmarks. We will give a quantitative comparison with that method in Section~\ref{Experiment}.
In~\cite{Tannu2019},  a variation-aware qubit movement strategy is proposed. It takes advantage of the change in error rate and a change-aware quantum circuit transformation strategy by trying to select the route with the lowest probability of failure. This strategy uses the error rate of  SWAPs to allocate logical qubits to physical ones, thus avoiding paths with high error rates as much as possible. In \cite{Lao2019} Lao et al. have shown that the fidelity of a circuit is related to the delay and the number of gates. Now some heuristic methods are also applied to other platforms such as  Surface-17~\cite{guerreschi2019,Lao2019}.
\end{itemize}

A heuristic search algorithm often uses an evaluation function to obtain an optimal solution. Existing solutions  mainly aim at inserting as few SWAP gates as possible \cite{Zulehner2017,2020Zhu,Li2018,Cowtan2019,2020Qubit} or using the fidelity of the generated circuit as the objective function~\cite{Tannu2019} or minimizing the overall circuit latency~\cite{Lao2019}. We provide three options of evaluation functions, considering the number of additional SWAP gates, the depth of the output circuit, or configuration checking with aspiration. We also equip evaluation functions with a look-ahead parameter.
At present, there are a number of methods that exploit the look-ahead idea, such as \cite{Shafaei2013,Zulehner2017,2020Zhu,Li2018}. In particular, the method of \cite{2020Zhu} can dynamically adjust the number of look-ahead gates. 
Inspired by it, we allow a dynamic look-ahead parameter and adjust the parameter according to the number of layers of the input circuit, but mainly focusing on  the nearest gates. 

\section{Preliminary}
\label{Background}
In this section, we introduce some notions and notations of quantum computing.

Classical information is stored in bits, while quantum information is stored in qubits. 
Besides two basic states $\ket{0}$ and $\ket{1}$,
a qubit can be in any linear superposition state like $\ket{\phi}=a\ket{0}+b\ket{1}$,
where $a,b\in \mathbb{C}$ satisfy the condition $|a|^{2}+|b|^{2}=1$.
The intuition is that $\ket{\phi}$ is in the state $\ket{0}$ with  probability $|a|^{2}$ and in the state $\ket{1}$ with probability $|b|^{2}$.
We use the letter $\textsf{q}$ (resp. $\textit{q}$) to denote a physical qubit (resp. logical qubit).

\begin{figure}[htbp] 			
	\centerline{ 
		\Qcircuit @C=0.5em @R=0.4em {
			\lstick{\textit{q}_\textit{0}} &  \qswap  				&    \rstick{\textit{q}_\textit{1}} \qw &&&&  \lstick{\textit{q}_\textit{0}} 	&  \ctrl{2}  		&  \targ  		&  \ctrl{2}  		&    \rstick{\textit{q}_\textit{1}} \qw &&&&  \lstick{\textit{q}_\textit{0}}  &  \ctrl{2}  		&   \gate{H}  		&\ctrl{2} 			&\gate{H}     	&\ctrl{2}			&    \rstick{\textit{q}_\textit{1}}\qw  \\
			&		\qwx	&&&\push{\rule{.3em}{0em}=\rule{.3em}{0em}}&		&  	&					&			
			&		&      	& 		&	\push{\rule{.3em}{0em}=\rule{.3em}{0em}}					&					&				&					&         			&&&&			 \\
			\lstick{\textit{q}_\textit{1}} &   \qswap\qwx	   		&     \rstick{\textit{q}_\textit{0}}  \qw &&&&    \lstick{\textit{q}_\textit{1}} 	&   \targ      		&  \ctrl{-2}    &   \targ      		&     \rstick{\textit{q}_\textit{0}}  \qw   &&&&  \lstick{\textit{q}_\textit{1}}  &   \targ      		&   \gate{H}      	&   \targ      		&\gate{H} 		&\targ      		&    \rstick{\textit{q}_\textit{0}}\qw 	   \\	 
			&			&&&&		&  	&					&					&					&       		& 					&						&					&				&					&         			&&&&			 
		}
	}
	\caption{Implementing a SWAP gate by using CNOT gates and H gates}
	\label{f:Decomposition}
\end{figure}
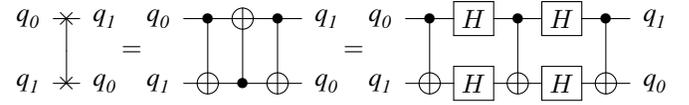

A quantum gate acts on a qubit to change the state of the qubit. For example, the Hadamard gate (H gate) is applied on a qubit, and the CNOT gate is applied on two qubits. Their symbols and matrix forms are shown in  Fig.~\ref{common_gates}.
We use a SWAP gate to exchange the states between two adjacent qubits, and multiple operations simulate moving non-adjacent qubits to adjacent positions.
A SWAP gate can be implemented by  three CNOT gates, or inserting  four H gates to change the direction of the middle CNOT gate, as shown in Fig.~\ref{f:Decomposition}. 
{
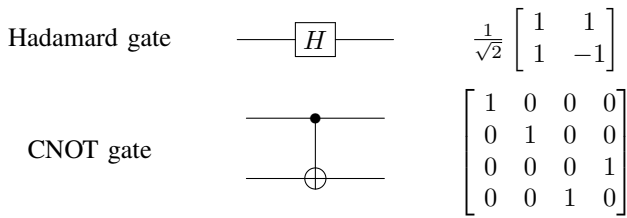
\begin{figure}[htbp]
	 \scalefont{1.0}
	 \begin{center}
	\begin{tikzpicture}
	\node at (7,0){	CNOT gate };
	\node at (10,0){	
		\Qcircuit @C=2.2em @R=1.75em {
		 & \ctrl{1}  & \qw 	 \\
		 &\targ  	 & \qw   \\	    			}};
	\node at (13,0){\vspace{10em}
				$\begin{bmatrix}
					\ 1\ & 0\ & 0\ & 0\ \\
					\ 0\ & 1\ & 0\ & 0\ \\
					\ 0\ & 0\ & 0\ & 1\ \\
					\ 0\ & 0\ & 1\ & 0\ 
				\end{bmatrix}$
	};



	\node at (7,1.5){	Hadamard gate };
	\node at (10,1.5){	\Qcircuit @C=2.2em @R=1.75em {
		 & \gate{H}  & \qw 	 \\  						 
	}};
	\node at (13,1.5){\vspace{10em}
				$\frac{1}{\sqrt{2}}\begin{bmatrix}
					\ 1\ & 1\ \\
					\ 1\ & -1\ 
				\end{bmatrix}$
	};
\end{tikzpicture}
\end{center}
\caption{The symbols of two quantum gates and their matrices}\label{common_gates}
\end{figure}	 

}

In a quantum circuit each line represents a \textit{wire}. The wire does not necessarily correspond to a physical wire, but may correspond to the passage of time or a physical particle that moves from one location to another through space. The interested reader can find more details of these gates from many textbooks such as \cite{Nielsen2016}.
The execution order of a quantum logical circuit  is from left to right.
The width 
of a circuit refers to the number of qubits in the circuit.
The depth 
of a circuit refers to the number of layers executable in parallel.
For example, the depth of the circuit in Fig.~\ref{OriginalCircuit} (a) is 6, and the width is 5.
We call a circuit with depth no more than 100 as a small-scale circuit, a circuit with depth more than 1000 as a large-scale circuit, 
and the rest are medium-scale circuits.
It is unnecessary to consider quantum gates acting on a single qubits in circuit adjustments, since 1-qubit gates are \textit{local}~\cite{Shafaei2013}.

\begin{figure}[htbp] 
	\begin{center}
		  {\scalefont{1.0}
	\begin{tikzpicture}
		\node at (5.5,3){(a) };
		\node at (5.5,5){  \Qcircuit @C=1.2em @R=0.75em {
			\lstick{\textit{q}_\textit{0}}   &  \qw 				&   \qw  \barrier{4}&\targ 	\barrier{4}	&\qw      		&\ctrl{4}\barrier{4}&   \qw		&\targ\barrier{4} &\targ  \barrier{4}&\qw  &  \qw       \\
			\lstick{\textit{q}_\textit{1}}   &   \targ      		&   \qw      		&   \qw      		&   \ctrl{2} 	&   \qw      	&   \targ    	&   \qw      	&   \qw       	&   \qw   		&  \qw       \\
			\lstick{\textit{q}_\textit{2}}   &   \ctrl{-1}  		&   \qw      		&   \qw      		&   \qw      	&   \qw      	&   \qw      	&   \qw     	&   \ctrl{-2} 	&   \targ       &  \qw       \\
			\lstick{\textit{q}_\textit{3}}   &\qw					&   \ctrl{1}   		&   \ctrl{-3} 		&   \targ    	&   \qw      	&   \ctrl{-2}	&   \qw      	&   \qw       	&   \qw        	&  \qw         \\
			\lstick{\textit{q}_\textit{4}}   &		\qw				& \targ				& \qw 				&   \qw      	&   \targ    	&   \qw      	&   \ctrl{-4}  	& 	\qw     	&   \ctrl{-2} 	&   \qw			\\
							  &\dstick{g_{0}}		&\dstick{g_{1}}		&\dstick{g_{2}}		&\dstick{g_{3}}	&\dstick{g_{4}} &\dstick{g_{5}} &\dstick{g_{6}} &\dstick{g_{7}} &\dstick{g_{8}}	&   		\\		 
							 &						&					&				&       		& 				& 				& 				&				&   				 
							 }};
	\end{tikzpicture}
		\begin{tikzpicture}
				\node at (5.5,5){  \Qcircuit @C=1.2em @R=0.75em {
				\lstick{\textit{q}_{\textit{0}}}   &   \qw  \barrier{4}&\targ 	&  \qw \barrier{4}	&\qw      		&\ctrl{4}\barrier{4}&   \qw		&\targ\barrier{4} &\targ  \barrier{4}&\qw  &  \qw       \\
				\lstick{\textit{q}_{\textit{1}}}    		&   \qw      		&   \qw      	&   \targ     	&   \ctrl{2} 	&   \qw      	&   \targ    	&   \qw      	&   \qw       	&   \qw   		&  \qw       \\
				\lstick{\textit{q}_{\textit{2}}}  		&   \qw      		&   \qw       &   \ctrl{-1}  		&   \qw      	&   \qw      	&   \qw      	&   \qw     	&   \ctrl{-2} 	&   \targ       &  \qw       \\
				\lstick{\textit{q}_{\textit{3}}}  			&   \ctrl{1}   		&   \ctrl{-3} 	 &\qw			&   \targ    	&   \qw      	&   \ctrl{-2}	&   \qw      	&   \qw       	&   \qw        	&  \qw         \\
				\lstick{\textit{q}_{\textit{4}}}   		& \targ				& \qw 	&		\qw					&   \qw      	&   \targ    	&   \qw      	&   \ctrl{-4}  	& 	\qw     	&   \ctrl{-2} 	&   \qw			\\
				&\dstick{g_{1}}		&\dstick{g_{2}}		&\dstick{g_{0}}		&\dstick{g_{3}}	&\dstick{g_{4}} &\dstick{g_{5}} &\dstick{g_{6}} &\dstick{g_{7}} &\dstick{g_{8}}	&   		\\		 
				&						&					&				&       		& 				& 				& 				&				&   				 
		}};
		\node at (5.5,3){(b) };
		\end{tikzpicture}
	}					 
	\caption{(a) The original quantum circuit. (b) The circuit after preprocessing.}
	\label{OriginalCircuit}	
	\end{center}
\end{figure}
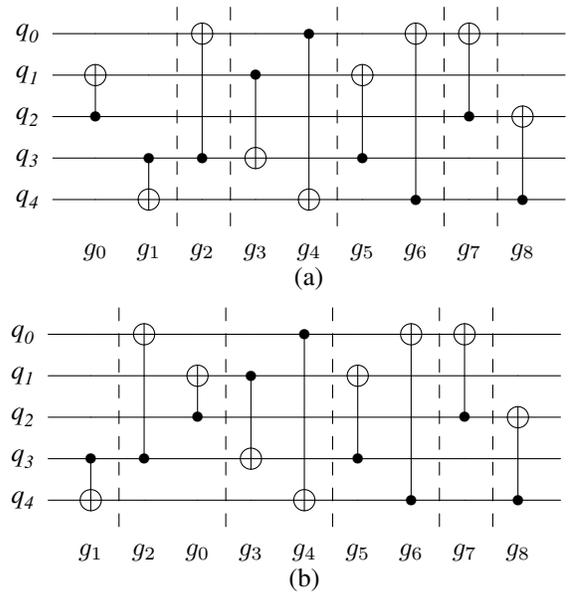


\begin{figure}
	{
  \scalefont{0.8}
  \begin{center}
  		\begin{tikzpicture}
  		\node at (1.8,0.4){(a) IBM QX2};
  		\node at (0.9,2.7){$\textsf{q}_\textsf{0}$};
  		\draw [black, thin] (0.9,2.7) circle [radius=0.25];
  		\node at (2.7,2.7){$\textsf{q}_\textsf{1}$};
  		\draw [black, thin] (2.7,2.7) circle [radius=0.25];
  		\node at (1.8,1.8){$\textsf{q}_\textsf{2}$};
  		\draw [black, thin] (1.8,1.8)circle [radius=0.25];
  		\node at (0.9,0.9){$\textsf{q}_\textsf{3}$};
  		\draw [black, thin] (0.9,0.9) circle [radius=0.25];
  		\node at (2.7,0.9){$\textsf{q}_\textsf{4}$};
  		\draw [black, thin] (2.7,0.9) circle [radius=0.25];
  		\draw [->,thin] (1.15,2.7) -- (2.45,2.7);
  		
  		\draw [<-,thin] (1.15,0.9) -- (2.45,0.9);
  		
  		\draw [->,thin] (1.075,2.525) -- (1.625,1.975);
  		\draw [->,thin] (2.525,2.525) -- (1.975,1.975);
  		\draw [->,thin] (1.075,1.075) -- (1.625,1.625);
  		\draw [->,thin] (2.525,1.075) -- (1.975,1.625);
  	\end{tikzpicture}
  \qquad
  	\begin{tikzpicture}
  		\node at (1.8,0.4){(b) IBM QX4};
  		\node at (0.9,2.7){$\textsf{q}_\textsf{0}$};
  		\draw [black, thin] (0.9,2.7) circle [radius=0.25];
  		\node at (2.7,2.7){$\textsf{q}_\textsf{1}$};
  		\draw [black, thin] (2.7,2.7) circle [radius=0.25];
  		\node at (1.8,1.8){$\textsf{q}_\textsf{2}$};
  		\draw [black, thin] (1.8,1.8)circle [radius=0.25];
  		\node at (0.9,0.9){$\textsf{q}_\textsf{3}$};
  		\draw [black, thin] (0.9,0.9) circle [radius=0.25];
  		\node at (2.7,0.9){$\textsf{q}_\textsf{4}$};
  		\draw [black, thin] (2.7,0.9) circle [radius=0.25];

  		\draw [<-,thin] (1.15,2.7) -- (2.45,2.7);
  		
  		\draw [<-,thin] (1.15,0.9) -- (2.45,0.9);
  		
  		\draw [<-,thin] (1.075,2.525) -- (1.625,1.975);
  		\draw [<-,thin] (2.525,2.525) -- (1.975,1.975);
  		\draw [<-,thin] (1.075,1.075) -- (1.625,1.625);
  		\draw [->,thin] (2.525,1.075) -- (1.975,1.625);
  	\end{tikzpicture}
  \end{center}
  \begin{center}
  	 \begin{tikzpicture}
  		\node at (3.15,1.7){(c) IBM QX3};
  		\draw [black, thin] (0,2.3) circle [radius=0.25];
  		\draw [<-,thin] (0.25,2.3) -- (0.65,2.3);
  		\draw [black, thin] (0.9,2.3) circle [radius=0.25];
  		\draw [->,thin] (1.15,2.3) -- (1.55,2.3);
  		\draw [black, thin] (1.8,2.3) circle [radius=0.25];
  		\draw [<-,thin] (2.05,2.3) -- (2.45,2.3);
  		\draw [black, thin] (2.7,2.3) circle [radius=0.25];
  		\draw [<-,thin] (2.95,2.3) -- (3.35,2.3);
  		\draw [black, thin] (3.6,2.3) circle [radius=0.25];
  		\draw [->,thin] (3.85,2.3) -- (4.25,2.3);
  		\draw [black, thin] (4.5,2.3) circle [radius=0.25];
  		\draw [->,thin] (4.75,2.3) -- (5.15,2.3);
  		\draw [black, thin] (5.4,2.3) circle [radius=0.25];
  		\draw [<-,thin] (5.65,2.3) -- (6.05,2.3);
  		\draw [black, thin] (6.3,2.3) circle [radius=0.25];
  		
  		\draw [->,thin] (0,2.55) -- (0,2.95);
  		\draw [<-,thin] (1.8,2.55) -- (1.8,2.95);
  		\draw [->,thin] (2.7,2.55) -- (2.7,2.95);
  		\draw [->,thin] (3.6,2.55) -- (3.6,2.95);
  		\draw [<-,thin] (4.5,2.55) -- (4.5,2.95);
  		\draw [<-,thin] (5.4,2.55) -- (5.4,2.95);
  		\draw [->,thin] (6.3,2.55) -- (6.3,2.95);
  		\draw [black, thin] (0,3.2) circle [radius=0.25];
  		\draw [->,thin] (0.25,3.2) -- (0.65,3.2);
  		\draw [black, thin] (0.9,3.2) circle [radius=0.25];
  		\draw [->,thin] (1.15,3.2) -- (1.55,3.2);
  		\draw [black, thin] (1.8,3.2) circle [radius=0.25];
  		\draw [<-,thin] (2.05,3.2) -- (2.45,3.2);
  		\draw [black, thin] (2.7,3.2) circle [radius=0.25];
  		\draw [->,thin] (2.95,3.2) -- (3.35,3.2);
  		\draw [black, thin] (3.6,3.2) circle [radius=0.25];
  		
  		\draw [black, thin] (4.5,3.2) circle [radius=0.25];
  		\draw [->,thin] (4.75,3.2) -- (5.15,3.2);
  		\draw [black, thin] (5.4,3.2) circle [radius=0.25];
  		\draw [<-,thin] (5.65,3.2) -- (6.05,3.2);
  		\draw [black, thin] (6.3,3.2) circle [radius=0.25];
  		\node at (0,2.3){$\textsf{q}_\textsf{0}$};
  		\node at (0.9,2.3){$\textsf{q}_\textsf{15}$};
  		\node at (1.8,2.3){$\textsf{q}_\textsf{14}$};
  		\node at (2.7,2.3){$\textsf{q}_\textsf{13}$};
  		\node at (3.6,2.3){$\textsf{q}_\textsf{12}$};
  		\node at (4.5,2.3){$\textsf{q}_\textsf{11}$};
  		\node at (5.4,2.3){$\textsf{q}_\textsf{10}$};
  		\node at (6.3,2.3){$\textsf{q}_\textsf{9}$};
  		
  		\node at (0,3.2){$\textsf{q}_\textsf{1}$};
  		\node at (0.9,3.2){$\textsf{q}_\textsf{2}$};
  		\node at (1.8,3.2){$\textsf{q}_\textsf{3}$};
  		\node at (2.7,3.2){$\textsf{q}_\textsf{4}$};
  		\node at (3.6,3.2){$\textsf{q}_\textsf{5}$};
  		\node at (4.5,3.2){$\textsf{q}_\textsf{6}$};
  		\node at (5.4,3.2){$\textsf{q}_\textsf{7}$};
  		\node at (6.3,3.2){$\textsf{q}_\textsf{8}$};

  	\end{tikzpicture}
  \end{center}
\begin{center}
	\begin{tikzpicture}
	
	\node at (3.15,-0.1){(d) IBM QX5};
\node at (0,0.5){$\textsf{q}_\textsf{0}$};
\node at (0.9,0.5){$\textsf{q}_\textsf{15}$};
\node at (1.8,0.5){$\textsf{q}_\textsf{14}$};
\node at (2.7,0.5){$\textsf{q}_\textsf{13}$};
\node at (3.6,0.5){$\textsf{q}_\textsf{12}$};
\node at (4.5,0.5){$\textsf{q}_\textsf{11}$};
\node at (5.4,0.5){$\textsf{q}_\textsf{10}$};
\node at (6.3,0.5){$\textsf{q}_\textsf{9}$};

\node at (0,1.4){$\textsf{q}_\textsf{1}$};
\node at (0.9,1.4){$\textsf{q}_\textsf{2}$};
\node at (1.8,1.4){$\textsf{q}_\textsf{3}$};
\node at (2.7,1.4){$\textsf{q}_\textsf{4}$};
\node at (3.6,1.4){$\textsf{q}_\textsf{5}$};
\node at (4.5,1.4){$\textsf{q}_\textsf{6}$};
\node at (5.4,1.4){$\textsf{q}_\textsf{7}$};
\node at (6.3,1.4){$\textsf{q}_\textsf{8}$};

\draw [black, thin] (0,0.5) circle [radius=0.25];
\draw [<-,thin] (0.25,0.5) -- (0.65,0.5);
\draw [black, thin] (0.9,0.5) circle [radius=0.25];
\draw [->,thin] (1.15,0.5) -- (1.55,0.5);
\draw [black, thin] (1.8,0.5) circle [radius=0.25];
\draw [<-,thin] (2.05,0.5) -- (2.45,0.5);
\draw [black, thin] (2.7,0.5) circle [radius=0.25];
\draw [<-,thin] (2.95,0.5) -- (3.35,0.5);
\draw [black, thin] (3.6,0.5) circle [radius=0.25];
\draw [->,thin] (3.85,0.5) -- (4.25,0.5);
\draw [black, thin] (4.5,0.5) circle [radius=0.25];
\draw [->,thin] (4.75,0.5) -- (5.15,0.5);
\draw [black, thin] (5.4,0.5) circle [radius=0.25];
\draw [<-,thin] (5.65,0.5) -- (6.05,0.5);
\draw [black, thin] (6.3,0.5) circle [radius=0.25];

\draw [<-,thin] (0,0.75) -- (0,1.15);
\draw [->,thin] (0.9,0.75) -- (0.9,1.15);
\draw [<-,thin] (1.8,0.75) -- (1.8,1.15);
\draw [->,thin] (2.7,0.75) -- (2.7,1.15);
\draw [->,thin] (3.6,0.75) -- (3.6,1.15);
\draw [<-,thin] (4.5,0.75) -- (4.5,1.15);
\draw [<-,thin] (5.4,0.75) -- (5.4,1.15);
\draw [->,thin] (6.3,0.75) -- (6.3,1.15);
\draw [black, thin] (0,1.4) circle [radius=0.25];
\draw [->,thin] (0.25,1.4) -- (0.65,1.4);
\draw [black, thin] (0.9,1.4) circle [radius=0.25];
\draw [->,thin] (1.15,1.4) -- (1.55,1.4);
\draw [black, thin] (1.8,1.4) circle [radius=0.25];
\draw [->,thin] (2.05,1.4) -- (2.45,1.4);
\draw [black, thin] (2.7,1.4) circle [radius=0.25];
\draw [<-,thin] (2.95,1.4) -- (3.35,1.4);
\draw [black, thin] (3.6,1.4) circle [radius=0.25];
\draw [<-,thin] (3.85,1.4) -- (4.25,1.4);
\draw [black, thin] (4.5,1.4) circle [radius=0.25];
\draw [->,thin] (4.75,1.4) -- (5.15,1.4);
\draw [black, thin] (5.4,1.4) circle [radius=0.25];
\draw [<-,thin] (5.65,1.4) -- (6.05,1.4);
\draw [black, thin] (6.3,1.4) circle [radius=0.25];

	\end{tikzpicture}
\end{center}
 \begin{center}
   \begin{tikzpicture}
 	
 	\node at (1.8,-0.1){(e) IBM Q20};
 	\draw [black, thin] (0,0.5) circle [radius=0.25];
 	\draw [<->,thin] (0.25,0.5) -- (0.65,0.5);
 	\draw [black, thin] (0.9,0.5) circle [radius=0.25];
 	\draw [<->,thin] (1.15,0.5) -- (1.55,0.5);
 	\draw [black, thin] (1.8,0.5) circle [radius=0.25];
 	\draw [<->,thin] (2.05,0.5) -- (2.45,0.5);
 	\draw [black, thin] (2.7,0.5) circle [radius=0.25];
 	\draw [<->,thin] (2.95,0.5) -- (3.35,0.5);
 	\draw [black, thin] (3.6,0.5) circle [radius=0.25];
 	
 	\node at (0,0.5) {$\textsf{q}_\textsf{15}$};
 	\node at (0.9,0.5){$\textsf{q}_\textsf{16}$};
 	\node at (1.8,0.5){$\textsf{q}_\textsf{17}$};
 	\node at (2.7,0.5){$\textsf{q}_\textsf{18}$};
 	\node at (3.6,0.5){$\textsf{q}_\textsf{19}$};
 	\draw [<->,thin] (0,0.75) -- (0,1.15);
 	\draw [<->,thin] (0.9,0.75) -- (0.9,1.15);
 	\draw [<->,thin] (1.8,0.75) -- (1.8,1.15);
 	\draw [<->,thin] (2.7,0.75) -- (2.7,1.15);
 	\draw [<->,thin] (3.6,0.75) -- (3.6,1.15);
 	
 	\draw [black, thin] (0,1.4) circle [radius=0.25];
 	\draw [<->,thin] (0.25,1.4) -- (0.65,1.4);
 	\draw [black, thin] (0.9,1.4) circle [radius=0.25];
 	\draw [<->,thin] (1.15,1.4) -- (1.55,1.4);
 	\draw [black, thin] (1.8,1.4)circle [radius=0.25];
 	\draw [<->,thin] (2.05,1.4) -- (2.45,1.4);
 	\draw [black, thin] (2.7,1.4) circle [radius=0.25];
 	\draw [<->,thin] (2.95,1.4) -- (3.35,1.4);
 	\draw [black, thin] (3.6,1.4) circle [radius=0.25];
 	\node at (0,1.4) {$\textsf{q}_\textsf{10}$};
 	\node at (0.9,1.4){$\textsf{q}_\textsf{11}$};
 	\node at (1.8,1.4){$\textsf{q}_\textsf{12}$};
 	\node at (2.7,1.4){$\textsf{q}_\textsf{13}$};
 	\node at (3.6,1.4){$\textsf{q}_\textsf{14}$};
 	\draw [<->,thin] (0,1.65) -- (0,2.05);
 	\draw [<->,thin] (0.9,1.65)-- (0.9,2.05);
 	\draw [<->,thin] (1.8,1.65) -- (1.8,2.05);
 	\draw [<->,thin] (2.7,1.65) -- (2.7,2.05);
 	\draw [<->,thin] (3.6,1.65) -- (3.6,2.05);
 	
 	\draw [black, thin] (0,2.3) circle [radius=0.25];
 	\draw [<->,thin] (0.25,2.3) -- (0.65,2.3);
 	\draw [black, thin] (0.9,2.3) circle [radius=0.25];
 	\draw [<->,thin] (1.15,2.3) -- (1.55,2.3);
 	\draw [black, thin] (1.8,2.3)circle [radius=0.25];
 	\draw [<->,thin] (2.05,2.3) -- (2.45,2.3);
 	\draw [black, thin] (2.7,2.3) circle [radius=0.25];
 	\draw [<->,thin] (2.95,2.3) -- (3.35,2.3);
 	\draw [black, thin] (3.6,2.3) circle [radius=0.25];
 	\node at (0,2.3) {$\textsf{q}_\textsf{5}$};
 	\node at (0.9,2.3){$\textsf{q}_\textsf{6}$};
 	\node at (1.8,2.3){$\textsf{q}_\textsf{7}$};
 	\node at (2.7,2.3){$\textsf{q}_\textsf{8}$};
 	\node at (3.6,2.3){$\textsf{q}_\textsf{9}$};
 	\draw [<->,thin] (0,2.55) -- (0,2.95);
 	\draw [<->,thin] (0.9,2.55)-- (0.9,2.95);
 	\draw [<->,thin] (1.8,2.55) -- (1.8,2.95);
 	\draw [<->,thin] (2.7,2.55) -- (2.7,2.95);
 	\draw [<->,thin] (3.6,2.55) -- (3.6,2.95);
 	
 	\draw [black, thin] (0,3.2) circle [radius=0.25];
 	\draw [<->,thin] (0.25,3.2) -- (0.65,3.2);
 	\draw [black, thin] (0.9,3.2) circle [radius=0.25];
 	\draw [<->,thin] (1.15,3.2) -- (1.55,3.2);
 	\draw [black, thin] (1.8,3.2)circle [radius=0.25];
 	\draw [<->,thin] (2.05,3.2) -- (2.45,3.2);
 	\draw [black, thin] (2.7,3.2) circle [radius=0.25];
 	\draw [<->,thin] (2.95,3.2) -- (3.35,3.2);
 	\draw [black, thin] (3.6,3.2) circle [radius=0.25];
 	\node at (0,3.2) {$\textsf{q}_\textsf{0}$};
 	\node at (0.9,3.2){$\textsf{q}_\textsf{1}$};
 	\node at (1.8,3.2){$\textsf{q}_\textsf{2}$};
 	\node at (2.7,3.2){$\textsf{q}_\textsf{3}$};
 	\node at (3.6,3.2){$\textsf{q}_\textsf{4}$};
 	
 	\draw [<->,thin] (1.075,0.675) -- (1.625,1.225);
 	\draw [<->,thin] (1.075,1.225) -- (1.625,0.675);
 	\draw [<->,thin] (2.875,0.675) -- (3.425,1.225);
 	\draw [<->,thin] (2.875,1.225) -- (3.425,0.675);
 	\draw [<->,thin] (1.075,2.475) -- (1.625,3.025);
 	\draw [<->,thin] (1.075,3.025) -- (1.625,2.475);
 	\draw [<->,thin] (2.875,2.475) -- (3.425,3.025);
 	\draw [<->,thin] (2.875,3.025) -- (3.425,2.475);
 	\draw [<->,thin] (0.175,2.125) -- (0.725,1.575);
 	\draw [<->,thin] (0.175,1.575) -- (0.725,2.125);
 	\draw [<->,thin] (1.975,2.125) -- (2.525,1.575);
 	\draw [<->,thin] (1.975,1.575) -- (2.525,2.125);
 \end{tikzpicture}
 \end{center}
}
\caption{IBM QX coupling graphs}
\label{IBM}
\end{figure}
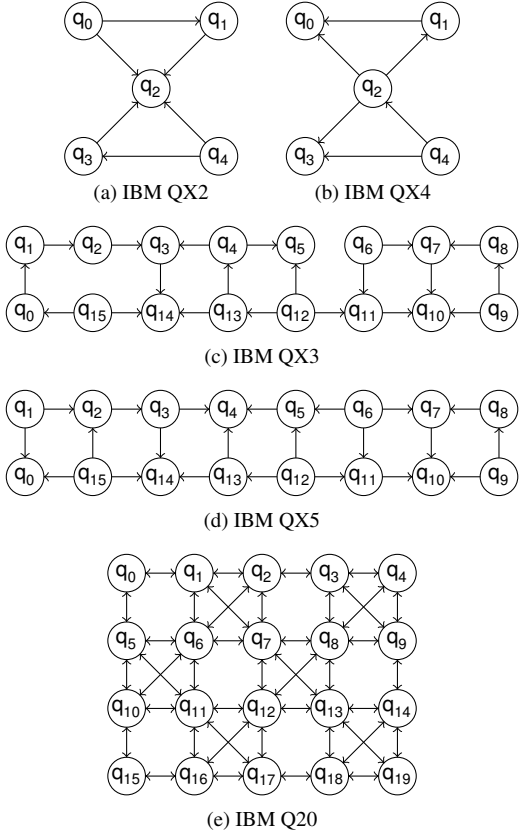

In the current work, we mainly consider the physical circuits of the IBM Q series, called coupling graphs.
Let $\mathcal{\mathcal{CG}}=(V_{\mathcal{C}}, E_{\mathcal{C}})$ denote the coupling graph of a physical device,
where $V_{\mathcal{C}}$ is the set of physical qubits and $E_{\mathcal{C}}$ is the set of edges representing the connectivity between qubits related by CNOT gates.
In Fig.~\ref{IBM}, (a) and (b) are the coupling graphs of the 5-qubit IBM QX2 and IBM QX4, respectively; (c) and (d) are the coupling graphs of the 16-qubit IBM QX3 and IBM QX5, respectively; and (e) is the coupling graph of the IBM Q20.
The direction in each edge indicates the control direction of each 2-qubit gate,
and 2-qubit gates can only be performed between two adjacent qubits.

For IBM QX2, QX3, QX4, and QX5, the control of one qubit to a neighbour is unilateral, but for IBM Q20 the control between two adjacent qubits is bilateral.

Assume an interaction graph $\mathcal{IG}$, a  coupling graph $\mathcal{CG}$, an initial mapping $\tau$, and a CNOT gate $g=\left \langle \textit{q}_\textit{i},\textit{q}_\textit{j}\right \rangle $, where $\textit{q}_\textit{i}$ is the control qubit and $\textit{q}_\textit{j}$ is the target qubit.
If the gate $g$ is executable on coupling graph $\mathcal{CG}$, then
$\left \langle\tau(\textit{q}_\textit{i}),\tau(\textit{q}_\textit{j})\right \rangle $ 
must be a directed edge on $\mathcal{CG}$.


\begin{figure}
	\begin{center}
		{\scalefont{1.0}
	\begin{tikzpicture}

		\node at (0.5,0){(a)};
            \draw [black, thin] (0,1.4) circle [radius=0.25];
            \draw [-,thin] (0.25,1.4) -- (0.65,1.4);
            \draw [black, thin] (0.9,1.4) circle [radius=0.25];
            \node at (0,1.4) {$\textit{q}_\textit{0}$};
            \node at (0.9,1.4){$\textit{q}_\textit{4}$};
            \draw [-,thin] (0,1.65) -- (0,2.05);
            \draw [-,thin] (0.9,1.65)-- (0.9,2.05);
        
            \draw [black, thin] (0,2.3) circle [radius=0.25];
        
            \draw [black, thin] (0.9,2.3) circle [radius=0.25];
            \node at (0,2.3) {$\textit{q}_\textit{3}$};
            \node at (0.9,2.3){$\textit{q}_\textit{2}$};
            \draw [-,thin] (0,2.55) -- (0,2.95);
        
            \draw [black, thin] (0,3.2) circle [radius=0.25];
            \draw [-,thin] (0.175,3.025) -- (0.725,2.475);
            \node at (0,3.2) {$\textit{q}_\textit{1}$};
			\draw [-,thin] (0.175,2.125) -- (0.725,1.575);
			\draw [-,thin] (0.175,1.575) -- (0.725,2.125);
			\label{LLAG}
			\end{tikzpicture}
			}
			\qquad   \qquad
			{
				\scalefont{0.8}
			\begin{tikzpicture}
            
            \node at (0.5,0){(b)};
            \draw [black, thin] (0,0.5) circle [radius=0.25];
            \draw [<->,thin] (0.25,0.5) -- (0.65,0.5);
            \draw [black, thin] (0.9,0.5) circle [radius=0.25];
        
            \node at (0,0.5) {$\textsf{q}_\textsf{15}$};
            \node at (0.9,0.5){$\textsf{q}_\textsf{16}$};
            \draw [<->,thin] (0,0.75) -- (0,1.15);
            \draw [<->,thin] (0.9,0.75) -- (0.9,1.15);
        
            \draw [black, thin] (0,1.4) circle [radius=0.25];
            \draw [<->,thin] (0.25,1.4) -- (0.65,1.4);
            \draw [black, thin] (0.9,1.4) circle [radius=0.25];
            \node at (0,1.4) {$\textsf{q}_\textsf{10}$};
            \node at (0.9,1.4){$\textsf{q}_\textsf{11}$};
            \draw [<->,thin] (0,1.65) -- (0,2.05);
            \draw [<->,thin] (0.9,1.65)-- (0.9,2.05);
        
            \draw [black, thin] (0,2.3) circle [radius=0.25];
            \draw [<->,thin] (0.25,2.3) -- (0.65,2.3);
            \draw [black, thin] (0.9,2.3) circle [radius=0.25];
            \node at (0,2.3) {$\textsf{q}_\textsf{5}$};
            \node at (0.9,2.3){$\textsf{q}_\textsf{6}$};
            \draw [<->,thin] (0,2.55) -- (0,2.95);
            \draw [<->,thin] (0.9,2.55)-- (0.9,2.95);
        
            \draw [black, thin] (0,3.2) circle [radius=0.25];
            \draw [<->,thin] (0.25,3.2) -- (0.65,3.2);
            \draw [black, thin] (0.9,3.2) circle [radius=0.25];
            \node at (0,3.2) {$\textsf{q}_\textsf{0}$};
            \node at (0.9,3.2){$\textsf{q}_\textsf{1}$};
        
        
            \draw [<->,thin] (0.175,2.125) -- (0.725,1.575);
			\draw [<->,thin] (0.175,1.575) -- (0.725,2.125);
			\label{PPAG}
	\end{tikzpicture}
			}
\end{center}
	
	\caption{(a) The logical interaction graph of the original circuit in Fig.~\ref{OriginalCircuit}(b). (b) The partial coupling graph of IBM Q20.}
	\label{LAGPAG}
\end{figure}
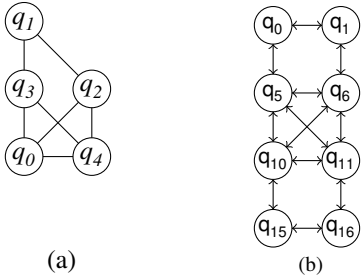

\begin{example}
	Suppose that a logical interaction graph $\mathcal{IG}$ and a coupling graph $\mathcal{CG}$ are shown in 	Fig.~\ref{LAGPAG}.
	Suppose that the initial mapping is as follows
	$$\tau=\{\textit{q}_\textit{0}\rightarrow  \textsf{q}_{\textsf{10}},\ \textit{q}_\textit{1}\rightarrow \textsf{q}_{\textsf{0}},\ 
	\textit{q}_\textit{2}\rightarrow  \textsf{q}_{\textsf{6}},\ \textit{q}_\textit{3}\rightarrow  \textsf{q}_{\textsf{5}},\ \textit{q}_\textit{4}\rightarrow  \textsf{q}_{\textsf{11}}\} .$$
Then the 2-qubit gate	$g_{0}=\left\langle \textit{q}_\textit{2},\textit{q}_\textit{1}\right\rangle$ is not executable, since the edge $\left \langle \tau(\textit{q}_\textit{2}),\tau(\textit{q}_\textit{1})\right \rangle =\left \langle \textsf{q}_{\textsf{6}},\textsf{q}_{\textsf{0}}\right \rangle $ does not exist in $\mathcal{CG}$.
However, the gate $g_{3}=\left\langle \textit{q}_\textit{1},\textit{q}_\textit{3}\right \rangle $ is executable, since 
	the edge $\left \langle \tau(\textit{q}_\textit{1}),\tau(\textit{q}_\textit{3})\right \rangle =\left \langle \textsf{q}_{\textsf{0}},\textsf{q}_{\textsf{5}}\right \rangle $  exists in $\mathcal{CG}$.
\end{example}

\section{Quantum Circuit Transformation}
\label{Quantum Circuit Transformation}
Assume that the input circuit has only 1-qubit gates and CNOT gates~\cite{1995Barenco,2005Mttnen}. We insert additional gates 
to move two non-adjacent qubits to adjacent positions or change the direction of CNOT gates. Inserting more gates increases the risk of introducing more noise. Therefore,
we expect to find a quantum circuit transformation algorithm that, when given an input circuit, can produce an output circuit with a small number of additional gates and a small depth in an acceptable amount of time.

Roughly speaking, we propose a method of quantum circuit transformation with the following three steps.

\leaveout{ 
\begin{figure}[htbp] 
	\centering
	\includegraphics[scale=0.4]{uml.jpg}		 
	\caption{Quantum Circuit transformation process}
	\label{processing}	
	 \end{figure}
 } 

\begin{enumerate}
	\item \emph{Preprocessing.} 
	This step includes extracting the interaction graph from the input circuit, shortening the lifetime of qubits as in~\cite{2019Zhang},  and calculating the shortest paths of the 
	coupling graph.
	\item \emph{Isomorphism and completion.} 
	This step first uses the subgraph isomorphism algorithm to find part of the initial mapping~\cite{Sun2020}.
	Then we perform a mapping completion to process the remaining nodes that do not satisfy all isomorphism requirements, according to the connectivity between the unmapped nodes and the mapped ones.
      \item \emph{Adjustment.} 
        After the second step, some logically adjacent nodes may be mapped to physically non-adjacent nodes, therefore, the quantum circuit is not executable on the coupling graph. It is necessary to adjust the quantum circuits by inserting additional gates. We use a tabu search-based adjustment algorithm to generate circuits that can be physically executed.
\end{enumerate}
Note that isomorphism and adjustment are both NP-complete~\cite{2018QubitSiraichi}. Thus, we make use of some heuristics. Below we give a detailed account of each step.

\subsection{Preprocessing}
In the preprocessing step, we adjust the input circuit described by an openQASM program~\cite{Andrew2017} to shorten the lifetime of qubits. Then we use a BFS search  to calculate the shortest distance between each pair of nodes on the coupling graph.

We use a layered method to analyze the lifetime of qubits and pack the gates that can be executed in parallel into a $bundle$, forming a layered bundle format~\cite{2019Zhang}.
Quantum gates acting on different qubits can be executed in parallel. Therefore, we classify the gates that can be executed in parallel into one layer; otherwise, we insert a new layer. The notation $\mathcal{L}(\mathcal{C}_{l})=\{\mathcal{L}_{0},\mathcal{L}_{1},...,\mathcal{L}_{n}\}$ denotes the layered circuit, where $\mathcal{L}_{i} \ (0 \le i \le n) $ stands for a quantum gate set that can be executed in parallel. The quantum gate set separated by the dotted  lines in Fig.~\ref{OriginalCircuit}(b) are the following: $\mathcal{L}_{0}=\{g_{1}\},\mathcal{L}_{1}=\{g_{2},g_{0}\},
\mathcal{L}_{2}=\{g_{3},g_{4}\},\mathcal{L}_{3}=\{g_{5},g_{6}\},\mathcal{L}_{4}=\{g_{7}\},\mathcal{L}_{5}=\{g_{8}\}$.

At the same time of layering, we generate an interaction graph $\mathcal{IG}=(V_{\mathcal{I}},E_{\mathcal{I}})$, which is an undirected graph with $V_{\mathcal{I}}$ being the set of vertices, and $E_{\mathcal{I}}$  the set of undirected edges that denotes the connectivity between qubits related by CNOT gates.
Given a coupling graph and assume the distance of each edge is 1, we use the Floyd-Warshall algorithm \cite{Floyd62a} to calculate the shortest distance matrix, with $dist[i][j]$ denoting the shortest distance from $\textsf{q}_{\textsf{i}}$ to $\textsf{q}_{\textsf{j}}$. 

Consider a CNOT gate $g=\left \langle  \textit{q}_\textit{i},\textit{q}_\textit{j} \right \rangle $. If $q_i$ and $q_j$  are mapped to $\textsf{q}_{m}$ and $\textsf{q}_{n}$, respectively, then the cost of executing $g$ under the shortest path is denoted by
$cost_{g}=7 \times( dist[m][n]-1)$ on devices with unilateral control. For IBM Q20, the cost is $cost_{g}=3 \times( dist[m][n]-1)$. 

\begin{example}
	Consider the QX5 coupling graph (cf. Fig.~\ref{IBM} (d)). Given a CNOT gate $g=\left \langle  \textit{q}_\textit{1}, \textit{q}_\textit{2} \right \rangle $, with \ $\textit{q}_\textit{1}$ mapped to $\textsf{q}_{6}$ and $\textit{q}_\textit{2}$ mapped to $\textsf{q}_{\textsf{13}}$, the shortest distance between them  is $dist[6][13]=3$. There are 3 shortest paths of moving from $\textsf{q}_{\textsf{6}}$ to an adjacent position of 
$\textsf{q}_{\textsf{13}}$:
$\pi_{0}={\textsf{q}_{\textsf{6}}\rightarrow \textsf{q}_{\textsf{5}} \rightarrow \textsf{q}_{\textsf{4}} \rightarrow \textsf{q}_{\textsf{13}}}$,
$\pi_{1}={\textsf{q}_{\textsf{6}}\rightarrow \textsf{q}_{\textsf{5}} \rightarrow \textsf{q}_{\textsf{12}} \rightarrow \textsf{q}_{\textsf{13}}}$,
$\pi_{2}={\textsf{q}_{\textsf{6}}\rightarrow \textsf{q}_{\textsf{11}} \rightarrow \textsf{q}_{\textsf{12}} \rightarrow \textsf{q}_{\textsf{13}}}$.
Their costs are given by 
$cost_{\pi_{0}}=18,\ cost_{\pi_{1}}=14,$ and $ cost_{\pi_{2}}=14$, respectively. Here $cost_{\pi}$ stands for the cost of swapping the qubits $\textit{q}_\textit{i}$ and $\textit{q}_\textit{j}$ by  following the path $\pi$.
\end{example}


\subsection{Isomorphism and Completion}

Generally speaking, in a coupling graph, it is almost impossible to find a subgraph that exactly matches the interaction graph. We regard the mapping with the largest number of mapped nodes as a good partial mapping. IMSM compares various compositions of several state-of-the-art subgraph isomorphism algorithms.  
It shows that the best performance can be achieved by using filters and the sorting ideas of the GraphQL algorithm to process candidate nodes, and the local candidates calculation method LFTJ based on set-intersection to enumerate the results. Since IMSM cannot process disconnected graphs, we artificially create connected graphs by linking isolated nodes to the nodes with the largest degree in the interaction graph. 
\begin{algorithm}[htbp]
	\caption{Complete the initial mapping}  
	\LinesNumbered  
	\KwIn{
	$\mathcal{CG}$: A coupling graph;\\
	$\mathcal{IG}$: An interaction graph; \\ 
	$T$: A partial mapping set obtained by IMSM;  \\
	}
	\KwOut{$results$: A set of mapping relations between
	 $\mathcal{IG}$ and $\mathcal{CG};$}
	\textbf{Initialize} $results=\emptyset$;\\
	$n \leftarrow \max_{\tau \in T}|\{i: \tau[i]\ne -1, i \le \tau.length, \; i\in N \}|$\\
	\For{$\tau\in T$}{
		\If{$n=\tau.length$}{
			$Q\leftarrow \{i: \tau[i]= -1,\; i\leq \tau.length,\; i\in N \}$; /*an empty unmapped node queue*/ \\
			
			\While{$Q \ne \emptyset$}{
				$\textit{q} \leftarrow Q.poll()$;\\
				$M_P \leftarrow$ $\mathcal{CG}.adjacencyMatrix()$; 	\\
				$M_L \leftarrow$ $\mathcal{IG}.adjacencyMatrix()$;	\\
					$C \leftarrow C \cup \{ \textit{q}_\textit{m}:  M_L[\textit{q}][\textit{q}_\textit{m}]\neq 0 \}$; /*sorted by degree*/\\
				\While{$C \ne \emptyset$}{
					$\textit{q}_\textit{m} \leftarrow \tau[C[0]]$; \\
					$k \leftarrow 0$; \\
					$C \leftarrow C\backslash C[0]$; \\
					\While{$ k < M_P[\textit{q}_\textit{m}].length$}{
						\If{$(M_P[\textit{q}_\textit{m}][k]\, or\, M_P[k][\textit{q}_\textit{m}]\neq 0  $ $ and\ !\tau.contains(k))$}{
							 $\tau[\textit{q}] \leftarrow k$; \\
							 break;
						 }
						  $k \leftarrow k+1$;
					}
					\If{ $k\neq M_P[\textit{q}_\textit{m}].length$ }{
						break;
					}
				}
			}
			$results.add(\tau)$;\\
		}
	}
	return $results$;
	\label{algorithm_initial}
	\end{algorithm}
	
The input of Algorithm 1 
 is a coupling graph $\mathcal{CG}$, an interaction graph $\mathcal{IG}$, and a partial mappings set $T$. We initialize an empty queue $Q$, and mark $q$, $\textsf{q}$ with integers such as $q_1=1,q_2=2,\textsf{q}_\textsf{1}=1.$ 
Then we traverse a mapping $\tau$ and add the unmapped nodes to the queue $Q$. For the unmapped nodes, we try to map them to the nodes adjacent to the mapped node in $\mathcal{CG}$. Finally, we generate a dense mapping. 
In principle, we could try to match the remaining unmapped nodes randomly, but it may lead to a mapping with a node far away from other nodes. If an unmapped node has an edge adjacent to a matched node in the interaction graph, it will first be matched to one of the adjacent nodes.  In this way, we can obtain all candidate mappings.

 In Algorithm 1,
 Line 2 selects the largest number $n$ of mapped nodes, and the partial mappings with $n$ mapped nodes are used by the candidate set.
 Lines 3-23 complete the partial mappings.
 In Line 5, we initialize an empty queue $Q$, which stores unmapped logical qubits, traverse the mapping $\tau$ and add the unmapped qubits to $Q$. We then loop until $Q$ is empty, and all logical qubits are mapped to physical qubits. We take out the first element in $Q$ to \textit{q}. Lines 8-9 get the adjacency matrices of $\mathcal{CG}$ and $\mathcal{IG}$, respectively. Line 10 initializes an empty map $C$, sorted by a descending order of the degree of
  connectivity between $\textit{q}_\textit{m}$ and \textit{q}.
 Lines 11-21 traverse  $C$, and select the node $\textit{q}_\textit{m}$ that has been mapped to the node $q$ in the coupling graph and  has the largest number of logical connections to \textit{q} in $C$.  Line 14 deletes the node from $C$. Lines 15-19 select the node $k$ adjacent to $q_{m}$ in the adjacency matrix, and map \textit{q} to that node.
\begin{example}
 Consider the interaction graph shown in Fig.~\ref{LAGPAG}~(a) and the coupling graph in Fig.~\ref{IBM}~(e). Suppose we have a partial mapping set $T=\{\tau_{0},\tau_{1},...,\tau_{n}\}$. We take one of the partial mappings as an example.
	$$\tau_{0}=\{\textit{q}_\textit{0}\rightarrow \textsf{q}_{\textsf{10}},\textit{q}_\textit{1}\rightarrow -1,
	\textit{q}_\textit{2}\rightarrow \textsf{q}_{\textsf{6}},\textit{q}_\textit{3}\rightarrow \textsf{q}_{\textsf{5}},\textit{q}_\textit{4}\rightarrow \textsf{q}_{\textsf{11}}\} $$ 
where $\textit{q}_\textit{1}\rightarrow -1$ means that $\textit{q}_\textit{1}$ is not mapped to any physical qubit, so we need  the mapping completion algorithm. The maximum number of mapped nodes is 4. We demonstrate how $\tau_{0}$ is completed. We add all unmapped nodes to the queue $Q$; in this example we have $Q=\{\textit{q}_\textit{1}\}$. Then we loop until $Q$ is empty. We pop the first element $q$ of $Q$, get the adjacency matrix of the query graph and the target graph, and traverse the adjacency matrix. We put the nodes  $\textit{q}_\textit{m}$ adjacent to \textit{q} into the candidate nodes list $C$, which is sorted by the connectivity of $\textit{q}_\textit{m}$ and \textit{q}. We get $C=\{\textit{q}_\textit{3},\textit{q}_\textit{2},\textit{q}_\textit{4},\textit{q}_\textit{0}\}$. Next, we traverse $C$ and take out  the first element $\textit{q}_\textit{3}$ in $C$, and calculate the physical node $q_{m}=\textsf{q}_{\textsf{5}}$  as
$\tau_0(\textit{q}_\textit{3})=\textsf{q}_{\textsf{5}}$. Finally, we map \textit{q} to the node connected to $q_{m}$ but not yet mapped. If the nodes connected to $q_{m}$ have been mapped, the loop continues. In this example, it can be directly mapped to $\textsf{q}_{\textsf{0}}$. In the end, we obtain the mapping $$ \tau_{0}=\{\textit{q}_\textit{0}\rightarrow  \textsf{q}_{10},\textit{q}_\textit{1}\rightarrow \textsf{q}_{0},	\textit{q}_\textit{2}\rightarrow  \textsf{q}_{\textsf{6}},\textit{q}_\textit{3}\rightarrow  \textsf{q}_{\textsf{5}},\textit{q}_\textit{4}\rightarrow  \textsf{q}_{\textsf{11}}\}. $$
	\end{example}
\subsection{Adjustment}
\subsubsection{Tabu search}
Tabu search is a type of heuristic algorithm. It uses a tabu list to avoid searching repeated spaces and  deadlock. The algorithm uses amnesty rules to jump out of  local optima to ensure the diversity of transformed results. Our circuit adjustment mainly relies on the tabu search algorithm, aiming to adjust those large-scale circuits that the existing algorithms are difficult to process and generate a circuit closer to the optimal solution.

The following objects are defined in tabu search: neighborhoods, neighborhood action, tabu list, candidate set, tabu object, evaluation function, and amnesty rule. All the edges that can be swapped are the neighborhoods. Obviously, it is not helpful to swap the edges that are not connected to  the mapped nodes, so we only add those edges on the shortest path to the candidate set. This does not affect the number of inserted additional gates, but saves significant amount of time rather than collect all the edges in the coupling graph into the candidate set. 
The tabu list avoids local optima and and tries to parallelize the inserted auxiliary gates. Tabu object is the object in the tabu list. We try not to use the recently swapped qubits as much as possible, which are added to the tabu list. Evaluation function selects an element from the candidate set that can make the nodes of the gate closer. Amnesty rules are used when all the objects in the candidate set are banned,  or after banning an object, the target value will be greatly increased.

The calculation of the candidate set is shown in Algorithm~2. 
The input $M_p$ contains the mapping from physical qubits to logical ones, where $ j = M_p[i] $ means that the $i$-th physical qubit  is mapped to the $j$-th logical qubit. The set $M_l $ denotes the mapping of logical qubits to physical qubits, where $j = M_l [i]$ means that the $i$-th logical qubit is mapped to the $j$-th physical qubit.
The set $L$ includes all the gates in the current layer, and the output is a candidate mapping set of the current mapping. The set $D$ contains the edges of all the shortest paths in the coupling graph. Lines 3-5 delete the gates that can be executed in $L$ under the current mapping. Lines 6-19 traverse gates $g$ in $L$, and calculate the shortest paths between the nodes of $g$. The edges involved in the shortest path are all the elements of the candidate set. Lines 9-15  update the mapping after the 
swap. Lines 16-19 generate a new candidate solution. Line 17 stores the swapped edges that will be used in the output circuit, and Line 18 calculates the swap scores.
\begin{algorithm} [htbp]
	\caption{Calculate the candidate set }  
	\LinesNumbered  
	\KwIn{
	$P$: The shortest paths set of coupling graph;\\
	$D$: The distance between nodes in the coupling graph;\\
	$M_p$: The mapping from physical qubits to logical qubits; \\
	$M_l$: The mapping from logical qubits to physical qubits; \\
	$L$: Gates included in the current layer of circuits; \\
	}
	\KwOut{$results$: The set of candidate mapping; }  
	\textbf{Initialize}  $results \leftarrow$ $\emptyset$;\\
					$ N \leftarrow $ An empty set of candidate swap nodes;\\  
					
					\ForEach{ $g\in  L$}{
						\If{$g \ is \ executable$}{
							$L \leftarrow L \backslash \{g\}$; \\
						}
					}
					
					\ForEach{ $g\in  L$}{
						\ForEach{$p \in P[M_l[g.control]][M_l[g.target]]$}{
						\ForEach{$e \in p $}{
							$M_q^{'} \leftarrow M_q$;
							$M_l^{'}  \leftarrow M_l$;\\
							$\textit{q}_\textit{1} \leftarrow M_q^{'}[e.s]$;
							$\textit{q}_\textit{2} \leftarrow M_q^{'}[e.t]$;\\
							$M_q^{'}[e.s]  \leftarrow  \textit{q}_\textit{2}$;
							$M_q^{'}[e.t]  \leftarrow  \textit{q}_\textit{1}$;\\
							\If{$\textit{q}_\textit{1}\neq -1$}{ 
								$M_l^{'}[\textit{q}_\textit{1}] \leftarrow  \textit{q}_\textit{2}$;
							}
							\If{$\textit{q}_\textit{2}\neq-1$}{
								$M_l^{'}[\textit{q}_\textit{2}] \leftarrow  \textit{q}_\textit{1}$;
							}
							$s \leftarrow \emptyset$;\\
							$s.swaps \leftarrow p.swaps \cup \{D[e.s][e.t]\};$\\
							$s.value  \leftarrow evaluate(D,M_l^{'},L)$;\\
							$results \leftarrow results \cup \{s\}$; \\
						
						}
					}
					}
          return $results$;
          
	\label{algorithm_neighborhood}
	\end{algorithm}

\begin{example}
	Let us consider the mapping $$\tau_{0}=\{\textit{q}_\textit{0}\rightarrow  \textsf{q}_{\textsf{10}},\textit{q}_\textit{1}\rightarrow \textsf{q}_{\textsf{0}},
\textit{q}_\textit{2}\rightarrow  \textsf{q}_{\textsf{6}},\textit{q}_\textit{3}\rightarrow  \textsf{q}_{\textsf{5}},\textit{q}_\textit{4}\rightarrow  \textsf{q}_{\textsf{11}}\} , $$ 
with  $L_{1}=\{g_{2},g_{0}\}$, $cost_{g_{2}}=3$ and $cost_{g_{0}}=6$.
The gate $g_{2}$ can be executed directly in the $\tau_{0}$ mapping, so we delete it from $L_{1}$,
but $g_{0}$ cannot be executed in the mapping $\tau_{0}$.
The nodes that cannot be executed join the set $swap\_nodes=\{\textsf{q}_{\textsf{0}},\textsf{q}_\textsf{6}\}$.
The set of shortest paths is $$paths=\{\{\textsf{q}_{\textsf{6}}\rightarrow \textsf{q}_{\textsf{1}} \rightarrow \textsf{q}_{0} \},\{\textsf{q}_\textsf{6}\rightarrow \textsf{q}_\textsf{5} \rightarrow \textsf{q}_\textsf{0} \}\},$$ 
We traverse the shortest paths to calculate the  candidate set.
The two endpoints of an edge passed by one of the shortest paths should intersect with the SWAP set and join the candidate set.
The current candidate set is $\{(\textsf{q}_\textsf{6},\textsf{q}_\textsf{1}),$ $(\textsf{q}_\textsf{1},\textsf{q}_\textsf{0}),$ $(\textsf{q}_\textsf{6},\textsf{q}_\textsf{5}),$ $(\textsf{q}_\textsf{5},\textsf{q}_\textsf{0}) \}$.
\end{example}

	\begin{algorithm}[htbp]
			\caption{Tabu search }  
			\LinesNumbered  
			\KwIn{$\tau_{ini}$: The initial mapping; \\
			$t_l$: Tabu list; 
			}
			\KwOut{$\tau_{best}$: The best mapping;}  
			\textbf{Initialize}
				$\tau_{best}  \leftarrow \tau_{ini}$; \\
				$iter \leftarrow 1$ /*the number of iterations*/ \\
			\While{$not \ mustStop(iter, \tau_{best})$}{
				$C \leftarrow \tau_{best}.candidates()$ /*candidate set*/ \\
				\If{C is empty}{
						$break$;
					}   
				$c_{best}  \leftarrow find\_best\_candidates(C, tl)$;\\
				\If{$c_{best}\ is\ empty$}{
						$c_{best} \leftarrow find\_amnesty\_candidates(C, tl)$;
				}
					$\tau_{best} \leftarrow c_{best}$;\\
				$t_l  \leftarrow t_l \cup\{c_{best}.swap\}$ ;\\
				$iter \leftarrow iter+1$;
			}
      return $\tau_{best}$
      
		\label{algorithm_Tabu}
	\end{algorithm}
TSA takes a layered circuit and an initial mapping as input and outputs a circuit that can be executed in the specified coupling graph, as shown in Algorithm~3. 
The adjusted circuit mapping of each layer is used as the initial mapping of the next layer. 
Line 1 regards the initial mapping $\tau_{ini}$ as the best mapping $\tau_{best}$. Lines 3-12 cyclically check whether all the gates in the current layer can be executed under the mapping $\tau_{best}$. If all the gates are executable or the number of iterations has reached the given bound, 
the search is completed. Otherwise, the search continues. Line 4 gets the current mapping candidate, and Line 7 finds the best mapping in the candidate set.
Note that if a SWAP appears in the tabu list, its corresponding mapping will be removed from the candidate set.  Then from the remaining candidates, we choose a mapping with the lowest cost. Line~9 takes the amnesty rules. If the best candidate is not found, the amnesty rules will select the mapping with the lowest cost in the candidate set as the best mapping. Lines 10-12 update the best mapping $\tau_{best}$, and insert the SWAP performed by the best mapping to the tabu list $t_l$, indicating that the SWAP in the tabu list should not be used as much as possible recently.
The algorithm would try to avoid re-swapping the just swapped qubits. Then it will check whether the termination condition of the algorithm is satisfied. The condition determines whether the number of iterations has reached the given bound, or the current mapping ensures all the gates in the current layer can be executed. 
\begin{example}
Let us continue the previous example. We select the one with the lowest evaluation scores from the candidate set.
For $\mathcal{L}_{1}=\{g_{2},g_{0}\}$, the candidate set is 
$\{(\textsf{q}_\textsf{6},\textsf{q}_\textsf{1}), (\textsf{q}_\textsf{1},\textsf{q}_\textsf{0}), (\textsf{q}_\textsf{6},\textsf{q}_\textsf{5}), (\textsf{q}_\textsf{5},\textsf{q}_\textsf{0}) \} , $ and the costs are given as follows.
\[\begin{array}{l}
cost_{num}(\textsf{q}_\textsf{6},\textsf{q}_\textsf{1})=3.0, \, cost_{num}(\textsf{q}_\textsf{1},\textsf{q}_\textsf{0})=3.0,\\ cost_{num}(\textsf{q}_\textsf{6},\textsf{q}_\textsf{5})=3.0, \, cost_{num}(\textsf{q}_\textsf{5},\textsf{q}_\textsf{0})=3.0 .
\end{array}\]

The algorithm chooses the first SWAP with the smallest score, and the mapping becomes $$\tau_{0}=\{\textit{q}_\textit{0}\rightarrow  \textsf{q}_\textsf{10},\textit{q}_\textit{1}\rightarrow  \textsf{q}_\textsf{0},
\textit{q}_\textit{2}\rightarrow  \textsf{q}_\textsf{1},\textit{q}_\textit{3}\rightarrow  \textsf{q}_\textsf{5},\textit{q}_\textit{4}\rightarrow  \textsf{q}_\textsf{11}\} . $$ 
 It can be seen that the current mapping ensures that $g_{0}$ is executable. So we can continue to the next layer. 
\end{example}
\subsubsection{Evaluation functions}
Evaluation functions can control the search direction.
We propose three evaluation functions: one introduces CCA, one uses the number of additional gates in the generated circuit as an evaluation criterion as given in (\ref{cost_num}),  and the last one uses the depth of the generated circuit as an evaluation criterion as given in (\ref{cost_depth}). They give rise to three variants of TSA called TSA$_{cca}$, TSA$_{num}$, and TSA$_{dep}$, respectively.

\begin{align}
	cost_{num}(\textsf{q}_\textsf{m},\textsf{q}_\textsf{n}) &=\sum_{g \in L}(dist[\tau(g.control)][\tau(g.target)])
    \label{cost_num} \\
	cost_{depth}(\textsf{q}_\textsf{m},\textsf{q}_\textsf{n}) &= Depth(L) \label{cost_depth}
\end{align}
Here $cost_{num}(\textsf{q}_\textsf{m},\textsf{q}_\textsf{n})$ (resp. $cost_{depth}(\textsf{q}_\textsf{m},\textsf{q}_\textsf{n})$) denotes the distance (resp. depth) between two qubits of all the gates in the layer $L$ 
after swapping $\textsf{q}_\textsf{m}$ with $\textsf{q}_\textsf{n}$. 

 CCA is a heuristic method, mainly used for SAT problems. We apply the 
 idea of CCA to adjust circuits. 
 Let $submake$ represent the number of qubits for which two qubits are closer after a SWAP operation, and $subbreak$ represent the number of qubits for which two qubits are farther apart after a SWAP operation.
 We introduce $subscore=submake-subbreak$ into the evaluation function, and adjust the weight with Smooth Weight based Threshold (SWT) scheme~\cite{Cai2012}.
The application of SWT in our experiment is mainly to add 1 to the edge of swap. When the weight of an edge exceeds the threshold, the weight of all edges becomes $\rho*w(c_i)+(1-\rho)* \bar{w}$, where $\rho$ is the influence factor of the weight of edge $c_i$ in the adjustment, and $\bar{w}$ represents the average weight. Adjusting the $\rho$ and threshold parameters has limited effect on our experiments, especially on large-scale circuits, because these two parameters only ensure that the weight is 
 close to the threshold.

\subsubsection{Look ahead}
The output of the $i$-th layer, with $i$ smaller than the depth of the circuit $d$,  is used as the input of the $(i+1)$-th layer. Note that any SWAP operation in the $i$-th layer will affect the mapping of the $(i+1)$-th layer. If we only consider the gates in the current layer when choosing the SWAP gates, the SWAP only satisfies the requirement of the $i$-th layer, not necessarily the next layer. Therefore, we take the gates 
from the $i$-th to the $(i+l_a)$-th layer, with $i+l_a \le d$,  into consideration, where $l_a$ is the number of look-ahead layers. However, it is necessary to give a higher priority to the execution of the gates in the $i$-th layer, so we introduce an attenuation factor $\delta$, which controls the influence of the gates in the look-ahead layers. 
Our evaluation functions in (\ref{cost_num}) and  (\ref{cost_depth}) can be modified as
(\ref{cost_num_look}) and  (\ref{cost_depth_look}), respectively.

\begin{align}
& cost_{num}(\textsf{q}_\textsf{m},\textsf{q}_\textsf{n}) =\left(\sum_{g \in L_{i}}dist[\tau(g.control)][\tau(g.target)] \right) \notag\\
&+ \delta \times \left(\sum_{j=i}^{i+l_a}\sum_{g \in L_{j}}dist[\tau(g.control)][\tau(g.target)] \right) 
	\label{cost_num_look} \\
& cost_{depth}(\textsf{q}_\textsf{m},\textsf{q}_\textsf{n})
= Depth(L_{i})+\delta \times Depth\left(\sum_{j=i}^{i+l_a}L_{j}\right).
		\label{cost_depth_look}
\end{align}

\subsubsection{Complexity}
Given an interaction graph  $\mathcal{IG}=(V_{\mathcal{I}},E_{\mathcal{I}})$ and a coupling graph $\mathcal{CG}=(V_{\mathcal{C}},E_{\mathcal{C}})$, we assume that 
the depth of the circuit is $d$. 
Tabu search processes each layer one by one, and searches at most $d$ times. Starting from the initial mapping, we first delete the executable gates of the first layer under the initial mapping. Then, the edges of all the shortest paths of all the gates that are not executable in the current layer are added to the candidate set where at least one node is in the gate mapping. In the worst case, the length of the shortest path  is $(|E_{\mathcal{C}}|-1)$
as well as the size of the candidate set. Each SWAP will make the gates closer. In the worst case, the number of SWAPs is $(|E_{\mathcal{C}}|-1)^{|E_{\mathcal{C}}|-2}$, but our selection strategy will make the number of SWAPs significantly reduced. The time complexity in the worst case is O($d\times (|E_{\mathcal{C}}|-1)^{(|E_{\mathcal{C}}|-2)}$), and the space complexity is the size of our candidate set $(E_{\mathcal{C}}-1)$, which is in PSPACE.
\section{Experiments}
\label{Experiment}
We compare TSA with several state-of-the-art algorithms for quantum circuit transformation, namely GA~\cite{Zulehner2017}, SABRE~\cite{Li2018}, FiDLS~\cite{2020Qubit} and  DLH~\cite{2020Zhu}. Notice that other algorithms such as SAHS~\cite{Xiangzhen2020} and $t\ket{ket}$\cite{Cowtan2019} are not listed because it has been pointed out in ~\cite{2020Qubit} that FiDLS is superior to SAHS and the latter outperforms $t\ket{ket}$ as shown in \cite{Xiangzhen2020}. The implementation in Python is available at \url{https://github.com/Holly-Jiang/QCTSA}.
All the experiments are conducted on a Ubuntu machine with 2.2GHz CPU and 64G memory. We take the logarithm $log_{10}$ of both the x-axis and y-axis such that the experimental results are easy to observe. The 
time limit for each benchmark is one hour.
Since SABRE uses a random initial mapping, for every testcase we execute it
five  times, each with a different initial mapping for each benchmark and reports the best result out of the five attempts. Other algorithms are deterministic, so it suffices to run them only once.

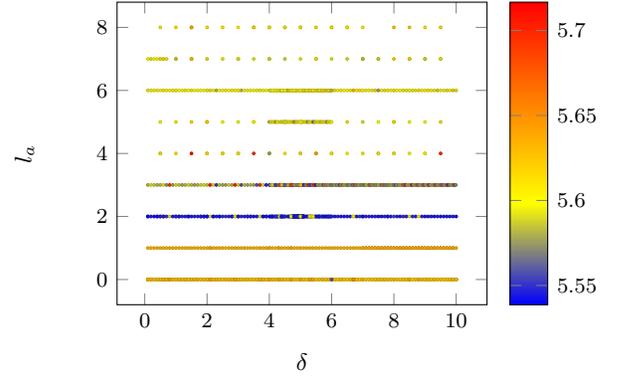
\begin{figure}[htbp] 			
	\centerline{ 
\begin{tikzpicture}
\begin{axis}[
colorbar,
xlabel=$\delta$, ylabel=$l_a$,
small,
]
\addplot+ [
scatter,
only marks,
mark size=0.5pt,
point meta={\thisrow{label}}
] table[x=x,y=y] {./charts/delta_forw.dat};
\end{axis}
\end{tikzpicture}
	}
 	\caption{The impact of the attenuation factor $\delta$ and the look-ahead parameter $l_a$ on search results. The x-axis represents $\delta$, the y-axis represents $l_a$, and the z-axis represents the number of additional gates.}
	\label{delta_forw}
\end{figure}

Firstly, 
we test TSA with fixed and variable look-ahead parameters $l_a$.
In Fig.~\ref{delta_forw}, different colors represent the logarithms 
of the number of additional gates.
The experiments show that the influence of look-ahead parameter $l_a$ is more significant than attenuation factor $\delta$.
The optimal attenuation factor for each circuit may be different. We have done thousands of experiments and 
found that when $\delta=0.5$ and $l_a=2$, the number of additional gates are relatively small for all benchmarks~\cite{Zulehner2017}. It means that the current layer is roughly twice as important as the later layers, and a 2-layer look-ahead already gives a good performance for TSA.

In Fig.~\ref{cca}, we compare  three evaluation functions TSA$_{cca}$, TSA$_{dep}$ and TSA$_{num}$. Using these indicators as objective functions, we test 159 benchmarks~\cite{Zulehner2017}. Compared with TSA$_{cca}$ (resp. TSA$_{dep}$), the depth of the generated circuits by TSA$_{num}$ is reduced by 4.02\% (resp. 3.24\%)  on average, and the number of additional gates are reduced by  26.22\% (resp. 24.52\%) on  average.  The evaluation function of TSA$_{cca}$ in quantum circuit transformation does not seem to have obvious advantage over TSA$_{num}$. Inserting a SWAP gate requires inserting 3 CNOT gates, and the depth will increase by 3. Therefore, inserting fewer SWAP gates may have a smaller circuit depth.
\begin{figure}[htbp] 			
	\centerline{
\begin{tikzpicture}
\begin{axis}[
enlargelimits=0,
legend style={at={(0.5,-0.14)},
anchor=north,legend columns=-1},
ybar,
bar width=1pt,
]
\addplot[color=blue,fill=blue] table [x=ini,y=num] {./charts/tsa_depth.dat};
\addplot[color=red,fill=red] table [x=ini,y=num] {./charts/cca.dat};
\addplot[color=green,fill=green] table [x=ini,y=num] {./charts/tsa.dat};
\legend{TSA$_{dep}$,TSA$_{cca}$,TSA$_{num}$}
\end{axis}
\end{tikzpicture}
	}
	\caption{Comparison of evaluation functions  TSA$_{dep}$, TSA$_{cca}$, and TSA$_{num}$. The x-axis represents the number of 2-qubit gates of the input circuit, and the y-axis represents the number of additional gates; similarly for the other figures in the rest of the article.}
	\label{cca}
\end{figure}
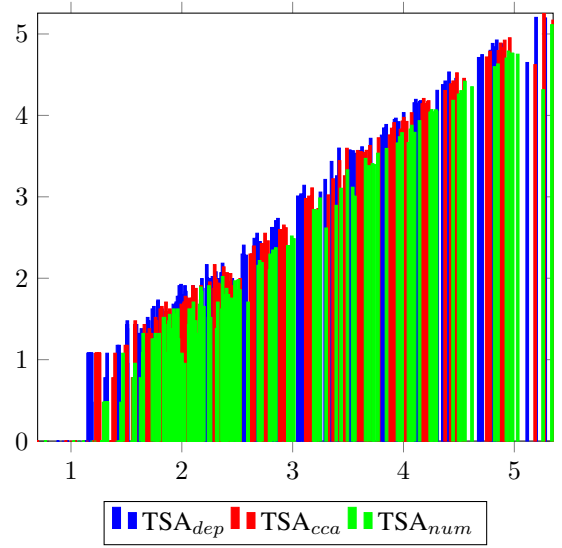
  \begin{table*}[htbp]
   \begin{center}
   \begin{threeparttable}
   \begin{tabular}{|p{1.8cm}<{\centering}|p{1cm}<{\centering}p{1cm}<{\centering}|p{1cm}<{\centering}|p{1.5cm}<{\centering}|p{1cm}<{\centering}|p{2cm}<{\centering}|p{2cm}<{\centering}|}
   \hline
   \multirow{2}*{benchmarks}&\multicolumn{2}{c|}{\#circ.}&MCPE &MCPE\_OP &TSA&  \multirow{2}*{$(g_0-g_2)/g_0$}& \multirow{2}*{$(g_1-g_2)/g_1$}\\
     \cline{2-6}		
   &$n$&$g$&$g_0$&$g_1$&$g_2$& & \\
   \hline
4mod5-v1\_22    & 5     & 21    &0      &0      & 0     &0\%        &0\%   \\
mod5mils\_65    & 5     & 35    &0      &0      &0      & 0\%       & 0\%    \\
alu-v0\_27      & 5     & 36    &3      & 3     &  6    & -100\%    & -100\%   \\
decod24-v2\_43  & 4     & 52    &0      & 0     &   0   & 0\%       & 0\%\\
4gt13\_92       &   5   & 66    &  21   &  21   &  0    &100\%      & 100\% \\
ising\_model\_16&16     & 786   & 0     & 0     &  0    &0\%        &0\%  \\
qft\_10         & 10    & 200   & 39    &  39   &  54   & -38.46\% &-38.46\%  \\
qft\_16         & 16    & 512   & 225   &  192  & 204   &9.33\% & -6.25\% \\
rd84\_142       & 15    & 343   & 153   &  108  & 93    &39.22\%  & 13.89\%  \\
radd\_250       & 13    & 3213  & 1353  & 1047  & 1086  &19.73\%  &-3.73\%  \\
z4\_268         &  11   & 3073  &  1071 & 855   & 747   &30.25\% &12.63\%   \\
sym6\_145       & 14    & 3888  &  1017 & 1017  &696   &31.56\% & 31.56\%   \\
misex1\_241     & 15    & 4813  &  2118 & 1098  & 1380 &34.84\% &-25.68\%    \\
rd73\_252       & 10    &  5321 &2352   &  2193 &1728   &26.53\% & 21.20\%  \\
cycle10\_2\_110 & 12    & 6050  & 2226  & 1968  & 2445  &-9.84\% & -24.24\%  \\
square\_root\_7 &  15   & 7630  & 2061  & 1788  & 1938 &5.97\% &-8.39\%  \\
sqn\_258        & 10    & 4459  & 3708  & 3057  & 3639 &1.86\% & -19.04\%  \\
rd84\_253       & 12    & 13658 & 6411  &5697   & 5061  &21.06\% &11.16\%  \\
co14\_215       &  15   & 17936 & 5634  &5062   & 7371 &-30.83\% & -45.61\%  \\
sym9\_193       & 10    & 34881 & 15420 &13746  &15318 &0.66\% &-11.43\%  \\
urf5\_158       & 9     & 164416 & 69852 &58947  &  71604  &-2.51\% &-21.47\%   \\
hwb9\_119       & 10    & 207775 &93219  &89355  & 63093 &32.32\% &29.39\%   \\
urf4\_187       & 11    & 512064 &220329 &168366 &  160728&27.05\% &4.54\%   \\
   \hline
sum             & 248   & 1307223&  427212 & 354558& 337191&21.07\% &4.90\%   \\
   \hline
   \end{tabular}
   \begin{tablenotes}
        \footnotesize
        \item[*]$n$: the number of qubits. $g$: the number of gates in the input circuit. $g_0$-$g_2$: the number of additional gates inserted by MCPE, MCPE\_OP and TSA, respectively.
   \end{tablenotes}
   \end{threeparttable}
   \end{center} 
   \caption{Comparison of MCPE, MCPE\_OP and TSA}
   \label{DLH}
   \end{table*}
\begin{figure}[htbp] 			
	\centerline{ 
\begin{tikzpicture}
\begin{axis}[
enlargelimits=0,
legend style={at={(0.5,-0.14)},
anchor=north,legend columns=-1},
ybar,
bar width=0.5pt,
]
\addplot[color=blue,fill=blue] table [color=green,x=ini,y=num] {./charts/optm.dat};
\addplot[color=red,fill=red] table [color=yellow,x=ini,y=num] {./charts/sabre_optm.dat};
\addplot[color=green,fill=green] table [color=yellow,x=ini,y=num] {./charts/FiDLS_optm.dat};
\addplot[color=yellow,fill=yellow] table [color=red,x=ini,y=num] {./charts/tsa_optm.dat};

\legend{\scalefont{0.5}GA,\scalefont{0.5}SABRE\_GA,\scalefont{0.5}FiDLS\_GA,\scalefont{0.5}TSA\_GA}
\end{axis}
\end{tikzpicture}
	}
	
	\caption{Comparison of the initial mapping algorithms of GA, SABRE, FiDLS and TSA, using the adjustment algorithm of GA. 
	}
	\label{f:initial_comparision_GA}
\end{figure}
\begin{figure}[htbp] 			
	\centerline{
\begin{tikzpicture}
\begin{axis}[
enlargelimits=0,
legend style={at={(0.5,-0.14)},
anchor=north,legend columns=-1},
ybar,
bar width=0.5pt,
]

\addplot[color=blue,fill=blue] table [color=black,x=ini,y=num] {./charts/optm_sabre.dat};
\addplot[color=red,fill=red] table [color=pink,x=ini,y=num] {./charts/FiDLS_sabre.dat};
\addplot[color=green,fill=green] table [color=yellow,x=ini,y=num] {./charts/SABRE.dat};
\addplot[color=yellow,fill=yellow] table [color=red,x=ini,y=num] {./charts/tsa_sabre.dat};

\legend{\scalefont{0.5}GA\_SABRE,\scalefont{0.5}SABRE,\scalefont{0.5}FiDLS\_SABRE,\scalefont{0.5}TSA\_SABRE}
\end{axis}
\end{tikzpicture}
	}

	\caption{Comparison of the initial mapping algorithms of GA, SABRE, FiDLS and TSA, using the adjustment algorithm of SABRE. 
	}
	\label{f:initial_comparision_sabre}
\end{figure}
\begin{figure}[htbp] 			
	\centerline{
\begin{tikzpicture}
\begin{axis}[
enlargelimits=0,
legend style={at={(0.5,-0.14)},
anchor=north,legend columns=-1},
ybar,
bar width=0.5pt,
]

\addplot[color=blue,fill=blue] table [color=black,x=ini,y=num] {./charts/optm_FiDLS.dat};
\addplot[color=red,fill=red] table [color=pink,x=ini,y=num] {./charts/FiDLS_FiDLS.dat};
\addplot[color=green,fill=green] table [color=yellow,x=ini,y=num] {./charts/sabre_FiDLS.dat};
\addplot[color=yellow,fill=yellow] table [color=red,x=ini,y=num] {./charts/tsa_FiDLS.dat};

\legend{\scalefont{0.5}GA\_FiDLS,\scalefont{0.5}SABRE\_FiDLS,\scalefont{0.5}FiDLS,\scalefont{0.5}TSA\_FiDLS}
\end{axis}
\end{tikzpicture}
	}

	\caption{Comparison of the initial mapping algorithms of GA, SABRE, FiDLS and TSA, using the adjustment algorithm of FiDLS. 
	}
	\label{f:initial_comparision_FiDLS}
\end{figure}
\begin{figure}[htbp] 			
	\centerline{ 
\begin{tikzpicture}
\begin{axis}[
enlargelimits=0,
legend style={at={(0.5,-0.14)},
anchor=north,legend columns=-1},
ybar,
bar width=0.5pt,
]

\addplot[color=blue,fill=blue] table [color=black,x=ini,y=num] {./charts/optm_tsa.dat};
\addplot[color=red,fill=red] table [color=yellow,x=ini,y=num] {./charts/sabre_tsa.dat};
\addplot[color=green,fill=green] table [color=red,x=ini,y=num] {./charts/FiDLS_tsa.dat};
\addplot[color=yellow,fill=yellow] table [color=pink,x=ini,y=num] {./charts/tsa.dat};

\legend{\scalefont{0.5}GA\_TSA,\scalefont{0.5}SABRE\_TSA,\scalefont{0.5}FiDLS\_TSA,\scalefont{0.5}TSA}
\end{axis}
\end{tikzpicture}
	}
	\caption{Comparison of the initial mapping algorithms of GA, SABRE, FiDLS and TSA, using the adjustment algorithm of TSA$_{num}$. 
	}
	\label{f:initial_comparision_num}
\end{figure}
\begin{figure}[htbp] 			
	\centerline{
\begin{tikzpicture}
\begin{axis}[
ybar,
enlargelimits=0,
legend style={at={(0.5,-0.14)},
anchor=north,legend columns=-1},
bar width=0.5pt,
]
\addplot[color=blue,fill=blue] table [x=ini,y=num] {./charts/optm_cca.dat};
\addplot[color=red,fill=red] table [x=ini,y=num] {./charts/sabre_cca.dat};
\addplot[color=green,fill=green] table [x=ini,y=num] {./charts/FiDLS_cca.dat};
\addplot[color=yellow,fill=yellow] table [x=ini,y=num] {./charts/tsa_cca.dat};

\legend{\scalefont{0.5}GA\_CCA,\scalefont{0.5}SABRE\_CCA,\scalefont{0.5}FiDLS\_CCA,\scalefont{0.5}TSA\_CCA}

\end{axis}

\end{tikzpicture}
	}
	\caption{Comparison of the initial mapping algorithms of GA, SABRE, FiDLS and TSA, using the adjustment algorithm of TSA$_{cca}$. 
	}
	\label{f:initial_comparision_cca}
\end{figure}
\begin{table*}[htbp]
	\scriptsize
	\begin{center} 
	\begin{threeparttable}
	\begin{tabular}{|p{1.2cm}<{\centering}|p{0.8cm}<{\centering}|p{0.8cm}<{\centering}|p{0.8cm}<{\centering}|p{0.8cm}<{\centering}|p{0.8cm}<{\centering}|p{0.8cm}<{\centering}|p{1.5cm}<{\centering}|p{1.5cm}<{\centering}|p{1.5cm}<{\centering}|}
	\hline
	    \diagbox[innerwidth=1.2cm]{y}{x} &$n$	&  $g$ & $g_0$ & $g_1$ &$g_2$ & $g_3 $&$(g_0-g_3)/g_0$  &$(g_1-g_3)/g_1$ & $(g_2-g_3)/g_2$\\
	\hline
	   GA 	&101 &16970  & 18309 & 14724 &12498 &8373 &54.27\%  & 43.13\%  & 33.01\%\\
	\hline
	   SABRE 	&  118  &37435 & 27804 & 27462 &24237 &16290 & 41.41\%  &  41.41\%  & 32.79\%\\
	\hline
	   FiDLS 	& 81 & 625 & 3063 & 1875 &1992 &1704 & 44.37\% & 9.12\% & 14.46\% \\
	\hline
	   TSA$_{num}$ 	&  120  &72830 & 49452 & 52254 &54075 &40935 & 17.22\%&  21.66\% &24.30\% \\
	\hline
	   TSA$_{cca}$ 	&  120 & 72830&57438& 76425  & 82902& 53073 &7.60\%  &  30.56\% &  35.98\%\\
	\hline
	\end{tabular} 
	   \begin{tablenotes}
        \footnotesize
        \item[*]  $n$: the number of circuits that all the four initial mapping algorithms can successfully transform. $g$: the number of 2-qubit gates in the input circuits.  $g_0$-$g_3$: the number of additional gates inserted by GA, SABRE, FiDLS and TSA, respectively. 
   \end{tablenotes}
   \end{threeparttable}
	\end{center} 
	\caption{Comparison of the initial mapping algorithms of GA, FiDLS, SABRE and TSA.}
	\label{ini_table}
	\end{table*}
\begin{table*}[htbp]
	\scriptsize
	\begin{center} 
	\begin{threeparttable}
	\begin{tabular}{|p{1.2cm}<{\centering}|p{0.8cm}<{\centering}|p{0.8cm}<{\centering}|p{0.8cm}<{\centering}|p{0.8cm}<{\centering}|p{0.8cm}<{\centering}|p{0.8cm}<{\centering}|p{0.8cm}<{\centering}|p{1.5cm}<{\centering}|p{1.5cm}<{\centering}|p{1.5cm}<{\centering}|p{1.5cm}<{\centering}|}
	\hline 
	    \diagbox[innerwidth=1.2cm]{y}{x}	&  $n$ & $g$& $g_0$ & $g_1$ &$g_2$ &$g_3$&$g_4$ &$(g_0-g_3)/g_0$&$(g_1-g_3)/g_1$  &$(g_2-g_3)/g_2$ &$(g_4-g_3)/g_4$ \\
	\hline
	   GA 	&94  &13031 & 14478 & 11055 &4938 & 7371&9090 & 49.09\% & 33.32\% & -49.27\% &18.91\% \\
	\hline
	   SABRE 	&  105 & 21711 & 19053 &16614 & 6204 &11898 & 16545 &37.55\%  &  28.39\% & -91.78\%& 28.09\%  \\
	\hline
	   FiDLS 	& 87 & 11001& 7665 & 9387 &3057 & 5052 & 9459& 34.09\% & 46.18\% & -65.26\%  &46.59\% \\
	\hline
	   TSA 	&  118 &31021& 19341 & 11262 & 8907 &12696 & 17733 &34.36\%  &  -12.73\% & -42.54\% & 28.41\% \\
	\hline
	\end{tabular} 
	   \begin{tablenotes}
        \footnotesize
        \item[*] $n$: the number of circuits that all the five adjustment algorithms can successfully transform. $g$: the number of 2-qubit gates in the input circuits. 
         $g_0$-$g_4$: the number of additional gates inserted by GA, SABRE, FiDLS, TSA$_{num}$ and TSA$_{cca}$, respectively.
   \end{tablenotes}
   \end{threeparttable}
	\end{center} 
	\caption{Comparison of the adjustment algorithms of GA, SABRE, FiDLS, TSA$_{num}$ and TSA$_{cca}$. }
	\label{adjust_table}
	\end{table*}
\begin{table*}[htbp]
	\scriptsize
   \begin{center}
   \begin{threeparttable}
   \begin{tabular}{|p{1.2cm}<{\centering}|p{0.4cm}<{\centering}|p{0.9cm}<{\centering}|p{0.4cm}<{\centering}|p{0.9cm}<{\centering}|p{1.1cm}<{\centering}|p{0.4cm}<{\centering}|p{0.6cm}<{\centering}|p{1.1cm}<{\centering}|p{0.4cm}<{\centering}|p{1.0cm}<{\centering}|p{0.9cm}<{\centering}|p{0.4cm}<{\centering}|p{1.0cm}<{\centering}|p{0.7cm}<{\centering}|}
   \hline
   \multirow{2}*{benchmarks}&\multirow{2}*{$n$}&\multirow{2}*{$g$}& \multicolumn{3}{c|}{SABRE} & \multicolumn{3}{c|}{FiDLS} &\multicolumn{3}{c|}{TSA$_{num}$}&\multicolumn{3}{c|}{TSA$_{cca}$} \\
     \cline{4-15}		
   &&&$n_0$&$g_0$&$t_0$&$n_1$&$g_1$&$t_1$&$n_2$&$g_2$&$t_2$&$n_3$&$g_3$&$t_3$\\
   \hline
   small&66&2614&66&2301 &1.38&66&687 &739.96   &66&705 & 12.12&66& 915& 16.66\\
   \hline
   medium&49&13330&49&10218&21.78&21&978 &410.74    &49& 4140&47.61&49&5874 & 66.35\\
   \hline
   large&44&1427641&29&162522&12409.63&6&3714  &7924.71    &44&868638 &1614.37&44&1177173 &2229.05\\
   \hline
   sum&159 &1443585&144&175041&12432.79&93&5379 &9075.41    &159&873483 &1674.10&159&1183962 &2312.06\\
   \hline
   \end{tabular}
     \begin{tablenotes}
        \footnotesize
        \item[*] $n$: the number of test circuits. $g$: the number of 2-qubit gates in the input circuits. $n_0$-$n_3$: the number of circuits successfully transformed by SABRE, FiDLS, TSA$_{num}$ and TSA$_{cca}$, respectively. $t_0$-$t_3$: runtime of SABRE, FiDLS, TSA$_{num}$ and  TSA$_{cca}$, respectively, in seconds.
        $g_0$-$g_3$: the number of additional gates inserted by SABRE, FiDLS, TSA$_{num}$ and TSA$_{cca}$, respectively.
   \end{tablenotes}
   \end{threeparttable}
   \end{center} 
   \caption{Comparison of runtime and the number of circuits successfully transformed by SABRE, FiDLS, TSA$_{num}$, TSA$_{cca}$, respectively.}
   \label{succ_number}
   \end{table*}
\begin{figure}[htbp] 			
	\centerline{ 
\begin{tikzpicture}
\begin{axis}[
enlargelimits=0,
legend style={at={(0.5,-0.14)},
anchor=north,legend columns=-1},
ybar,
bar width=0.5pt,
]
\addplot[color=red,fill=red] table [x=ini,y=num] {./charts/optm_cca.dat};
\addplot[color=blue,fill=blue] table [x=ini,y=num] 
{./charts/optm.dat};
\addplot[color=green,fill=green] table [x=ini,y=num] {./charts/optm_sabre.dat};
\addplot[color=olive,fill=olive] table [x=ini,y=num] {./charts/optm_tsa.dat};
\addplot[color=yellow,fill=yellow] table [x=ini,y=num] {./charts/optm_FiDLS.dat};
\legend{\scalefont{0.5}GA\_CCA,\scalefont{0.5}GA,\scalefont{0.5}GA\_SABRE,\scalefont{0.5}GA\_TSA,\scalefont{0.5}GA\_FiDLS}
\end{axis}
\end{tikzpicture}
	}
	\caption{Comparison of the adjustment algorithms of GA, SABRE, FiDLS, TSA$_{num}$ and TSA$_{cca}$, using the initial mapping algorithm of GA. 
	}
	\label{f:adjustment_comparision_ga}
\end{figure}
\begin{figure}[htbp] 			
	\centerline{ 
\begin{tikzpicture}
\begin{axis}[
enlargelimits=0,
legend style={at={(0.5,-0.14)},
anchor=north,legend columns=-1},
ybar,
bar width=0.5pt,
]

\addplot[color=red,fill=red]  table [color=red,x=ini,y=num]{./charts/SABRE.dat}; 
\addplot[color=blue,fill=blue] table [x=ini,y=num] {./charts/sabre_optm.dat};
\addplot[color=green,fill=green] table [color=red,x=ini,y=num]{./charts/sabre_cca.dat};
\addplot[color=olive,fill=olive] table [color=pink,x=ini,y=num]  {./charts/sabre_tsa.dat};
\addplot[color=yellow,fill=yellow]  table [color=red,x=ini,y=num]{./charts/sabre_FiDLS.dat}; 

\legend{\scalefont{0.5}SABRE,\scalefont{0.5}SABRE\_GA,\scalefont{0.5}SABRE\_CCA,\scalefont{0.5}SABRE\_TSA,\scalefont{0.5}SABRE\_FiDLS}
\end{axis}
\end{tikzpicture}
	}
	\caption{Comparison of the adjustment algorithms of GA, SABRE, FiDLS, TSA$_{num}$ and TSA$_{cca}$, using the initial mapping algorithm of SABRE. 
	}
	\label{f:adjustment_comparision_sabre}
\end{figure}
\begin{figure}[htbp] 			
	\centerline{ 
\begin{tikzpicture}
\begin{axis}[
enlargelimits=0,
legend style={at={(0.5,-0.14)},
anchor=north,legend columns=-1},
ybar,
bar width=0.5pt,
]

\addplot[color=red,fill=red]  table [color=red,x=ini,y=num]{./charts/FiDLS_sabre.dat}; 
\addplot[color=blue,fill=blue] table [color=pink,x=ini,y=num]  {./charts/FiDLS_tsa.dat};
\addplot[color=green,fill=green] table [color=red,x=ini,y=num]{./charts/FiDLS_cca.dat};
\addplot[color=olive,fill=olive] table [x=ini,y=num] {./charts/FiDLS_optm.dat};
\addplot[color=yellow,fill=yellow]  table [color=red,x=ini,y=num]{./charts/FiDLS_FiDLS.dat}; 

\legend{\scalefont{0.5}FiDLS\_SABRE,\scalefont{0.5}FiDLS\_TSA,\scalefont{0.5}FiDLS\_CCA,\scalefont{0.5}FiDLS\_GA,\scalefont{0.5}FiDLS}
\end{axis}
\end{tikzpicture}
	}
	\caption{Comparison of the adjustment algorithms of GA, SABRE, FiDLS, TSA$_{num}$ and TSA$_{cca}$, using the initial mapping algorithm of FiDLS. 
	}
	\label{f:adjustment_comparision_FiDLS}
\end{figure}
\begin{figure}[htbp] 			
	\centerline{ 
\begin{tikzpicture}
\begin{axis}[
enlargelimits=0,
legend style={at={(0.5,-0.14)},
anchor=north,legend columns=-1},
ybar,
bar width=0.5pt,
]

\addplot[color=olive,fill=olive] table [x=ini,y=num] {./charts/cca.dat};
\addplot[color=blue,fill=blue] table [color=red,x=ini,y=num] {./charts/tsa_optm.dat};
\addplot[color=red,fill=red]  table [color=red,x=ini,y=num]{./charts/tsa.dat}; 
\addplot[color=green,fill=green] table [color=pink,x=ini,y=num] {./charts/tsa_sabre.dat};
\addplot[color=yellow,fill=yellow] table [color=red,x=ini,y=num] {./charts/tsa_FiDLS.dat};

\legend{\scalefont{0.5}TSA\_CCA,\scalefont{0.5}TSA\_GA,\scalefont{0.5}TSA,\scalefont{0.5}TSA\_SABRE,\scalefont{0.5}TSA\_FiDLS}
\end{axis}
\end{tikzpicture}

	}
	\caption{Comparison of the adjustment algorithms of GA, SABRE, FiDLS, TSA$_{num}$ and TSA$_{cca}$, using the initial mapping algorithm of TSA. 
	}
	\label{f:adjustment_comparision_tsa}
\end{figure}
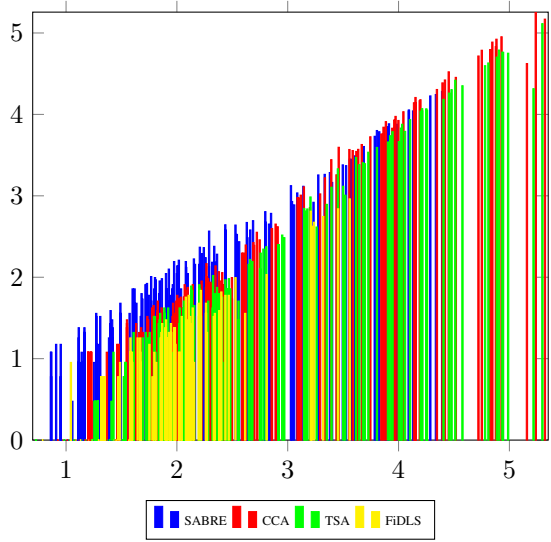
\begin{figure}[htbp] 			
	\centerline{
\begin{tikzpicture}
\begin{axis}[
enlargelimits=0,
legend style={at={(0.5,-0.14)},
anchor=north,legend columns=-1},
ybar,
bar width=0.5pt,
]
\addplot[color=blue,fill=blue]table [color=grey,x=ini,y=num] {./charts/SABRE.dat};
\addplot[color=red,fill=red]table [color=grey,x=ini,y=num] {./charts/tsa_cca.dat};
\addplot[color=green,fill=green] table [color=blue,draw=blue,fill=blue,x=ini,y=num] {./charts/tsa.dat};
\addplot[color=yellow,fill=yellow]table [color=green,x=ini,y=num] {./charts/FiDLS.dat};
\legend{\scalefont{0.5}SABRE, \scalefont{0.5}CCA, \scalefont{0.5}TSA, \scalefont{0.5}FiDLS}
\end{axis}
\end{tikzpicture}
	}
	\caption{Comparison of SABRE, FiDLS,TSA$_{num}$ and TSA$_{cca}$. 
	}
	\label{three_compa}
\end{figure}

Secondly, we compare TSA 
with  DLH, using the benchmarks in~\cite{2020Zhu}. Note that two evaluation functions MCPE and MCPE\_OP are used in DLH. Since there is no code for DLH available online, 
we only compare the number of additional gates inserted in the output circuits generated by DLH and TSA, as we can see in Table~\ref{DLH}.
Compared with MCPE and MCPE\_OP, TSA reduces the total number of additional gates by 21.07\%  and 4.90\%, respectively. 

Thirdly, we compare the combinations of several algorithms in the hope of inserting fewer additional gates.
We use the initial mapping and adjustment algorithms from GA~\cite{Zulehner2017}, SABRE~\cite{Li2018},  FiDLS~\cite{2020Qubit}, 
and TSA.

We compare the performance of the four initial mapping algorithms from GA, FiDLS, SABRE and TSA
under specific adjustment algorithms.
The five rows in Table~\ref{ini_table} correspond to Figs.~\ref{f:initial_comparision_GA}-\ref{f:initial_comparision_cca}.
For example, in the first row the adjustment algorithm is fixed to be that of GA; there are 101 circuits that all the four initial mapping algorithms can successfully transform and we compare the number of additional gates.
In Fig.~\ref{f:initial_comparision_GA}, 
the adjustment algorithm is fixed to be that of GA.
It can be seen that the initial mapping algorithm of TSA 
leads to a reduction of 54\%, 43\% and 33\% of additional gates than the initial mapping algorithms of GA, SABRE and FiDLS, respectively.
In Figs.~\ref{f:initial_comparision_sabre}-\ref{f:initial_comparision_cca}, 
the adjustment algorithm is fixed to be that of SABRE, FiDLS, TSA$_{num}$, and TSA$_{cca}$, respectively.
As can be seen from Table~\ref{ini_table}, on all the benchmarks, the initial mapping algorithm of TSA performs best when used in conjunction with the five adjustment algorithms.

In Figs.~\ref{f:adjustment_comparision_ga}-\ref{f:adjustment_comparision_tsa}, 
we compare the five adjustment algorithms  GA, SABRE, FiDLS, TSA$_{num}$ and TSA$_{cca}$ under specific initial mapping algorithms. For small-scale and medium-scale circuits FiDLS gives rise to the fewest additional gates. Since it is based on depth-first search, FiDLS takes large search space and long search time and the scale of circuits it can process is very limited. On the contrary, TSA performs well on large circuits in terms of additional gates and runtime.
The four rows in Table~\ref{adjust_table} correspond to Figs.~\ref{f:adjustment_comparision_ga}-\ref{f:adjustment_comparision_tsa}. For example, in the first row the initial mapping algorithm is fixed to be that of GA; there are 94 circuits that all the five adjustment algorithms can transform. In columns 4-8 we list the number of additional gates required by the five adjustment algorithms.

Finally, we compare the overall performance of TSA$_{num}$ and TSA$_{cca}$ with SABRE and FiDLS. We test 159 circuits, including 66 small-scale circuits, 49 medium-scale circuits and 44 large-scale ones. Note that in Table~\ref{succ_number} and Fig.~\ref{three_compa} we do not display the data for GA. Instead, we compare with SABRE because it is already shown in \cite{Li2018} that SABRE is much more scalable than GA. SABRE successfully transforms 144 circuits, including all the small-scale and medium-scale circuits, and 29 large-scale ones, which takes 12433 seconds. FiDLS successfully transforms 93 circuits, including all the small-scale circuits,  21 medium-scale circuits,  and 6 large-scale circuits, which takes 9075 seconds.  TSA$_{num}$ and TSA$_{cca}$ are much faster, as they successfully transform all the 159 circuits, which takes 1674 seconds and 2312 seconds, respectively.
Compared with SABRE, the number of additional SWAP gates generated by TSA$_{num}$ is reduced by  61\% on average, among the 115 small-scale and medium-scale circuits that both of them can successfully transform. More specifically, TSA$_{num}$ has  105  circuits with fewer additional gates than SABRE, and 9 circuits with equal numbers of gates.
Compared with TSA$_{num}$, the number of additional SWAP gates generated by FiDLS is reduced by 10\% on average, among the 87 small-scale and medium-scale circuits successfully transformed by it. Specifically, FiDLS inserts 555 additional gates, and for TSA$_{num}$ the number is 618. 
Although TSA$_{num}$ inserts a little more additional gates, it can transform large-scale circuits much more quickly, as we can see in Table~\ref{succ_number}. 

\section{Conclusion}
  \label{Conclusion}
  We proposed a scalable algorithm for quantum circuit transformation. We first used a subgraph isomorphism algorithm and a mapping completion algorithm based on the connectivity between qubits to generate a high-quality initial mapping. Then we employed a look-ahead heuristic search to adjust the mapping, which took into account the influence of the  gates yet to be processed to reduce the number of additional gates. 
  Compared with DLH, TSA performed better on large-scale circuits and took a shorter time in the DLH benchmarks.
  We compared the performance of the initial mapping and adjustment algorithm with the state-of-the-art algorithms GA, SABRE and FiDLS.
  Our experimental results showed that the initial mapping of TSA gave rise to fewer SWAP gates inserted and the adjustment algorithm could be obtained in an acceptable amount of time. Most small-scale and medium-scale circuits could be transformed in a few seconds.
  For large-scale circuits, the results could be obtained within a few minutes.
  In the future, we would investigate how to reduce the number of additional gates inserted and increase the speed. We would also apply the proposed method to more NISQ devices.

\bibliographystyle{IEEEtran}
\bibliography{mybibfile}

\end{document}